\documentclass[a4paper,11pt]{article}
\usepackage{jheppub} 
\usepackage{lineno}
\usepackage{orcidlink}
\usepackage{slashed}
\usepackage{subcaption}
\usepackage{esvect}
\graphicspath{{Figures/}}

\title{\boldmath Subleading Twist-3 Gluon Generalized Parton Distributions in the Light-Front Model}

\author[a]{Parashmani Thakuria\orcidlink{0000-0002-5652-7835}}
\author[b]{Madhurjya Lalung\orcidlink{0000-0002-7763-5050}}
\author[a]{Jayanta Kumar Sarma}

\affiliation[a]{Department of Physics, School of Sciences, Tezpur University, Tezpur, India, Pin-784028}
\affiliation[b]{Department of Physics, Nagaon University, Nagaon, India}

\emailAdd{parasht@tezu.ernet.in}
\emailAdd{mlalung2016@gmail.com}
\emailAdd{jks@tezu.ernet.in}

\abstract{We present a calculation of the twist-3 generalized parton distributions (GPDs) for gluons in the proton. Our analysis is performed within a light-front constituent model where the proton is treated as a two-body state of a spin-1 gluon and a spin-1/2 spectator system. The requisite light-front wave functions are derived from the soft-wall AdS/QCD correspondence. We compute the complete set of twist-3 gluon GPDs over a broad kinematic range. The corresponding distributions in impact parameter space are obtained via Fourier transform, revealing the transverse spatial distribution of gluons. Furthermore, we evaluate the contribution of these GPDs to the gluon kinetic orbital angular momentum (OAM) and compare our findings with other theoretical predictions.}


\begin{document}
\maketitle
\flushbottom

\section{Introduction}
\label{sec:intro}

One of the central goals of hadronic physics is to understand the internal structure of hadrons such as the proton and neutron—in terms of their fundamental constituents: quarks and gluons. For decades, this structure has been probed using parton distribution functions (PDFs), which describe the longitudinal momentum fraction carried by a parton inside a fast-moving hadron~\cite{deur2019spin, aidala2013spin, kumericki2016measuring}. PDFs are extracted from high-energy inclusive processes, such as deep inelastic scattering (DIS), where only the final-state lepton is detected. These distributions are forward matrix elements of bilocal light-cone operators and allow for a probabilistic interpretation of parton momentum distributions.
However, PDFs provide only a one-dimensional picture of hadrons. They encode information solely about the longitudinal momentum, with no access to the transverse spatial structure or parton correlations. To obtain a three-dimensional understanding of hadrons in momentum and coordinate space, a more general framework is required. This is achieved by generalized parton distributions (GPDs). GPDs~\cite{ji1997deeply, radyushkin1996scaling, diehl2003generalized, belitsky2002theory} extend PDFs by incorporating dependence on the squared momentum transfer \( t \) and longitudinal momentum transfer (skewness) \( \xi \). These are non-forward matrix elements of the same operators that define PDFs and encode rich correlations between partons, including their spatial distributions in the transverse plane.

GPDs reduce to PDFs in the forward limit \( \xi \to 0 \), \( t \to 0 \), and to form factors upon integration over the momentum fraction \( x \). Through their second Mellin moments, they are connected to the gravitational form factors (GFFs) of the nucleon, which describe the distribution of mass, pressure, and angular momentum within hadrons~\cite{polyakov2003gravitational, polyakov2018mechanical}. GPDs are experimentally accessed through exclusive processes such as deeply virtual Compton scattering (DVCS) and deeply virtual meson production (DVMP)~\cite{belitsky2001deeply, guidal2013exclusive, goloskokov2007longitudinal, goloskokov2011transversity}, and data from HERMES~\cite{airapetian2012beam}, COMPASS~\cite{d2004feasibility}, H1~\cite{adloff2002diffractive}, ZEUS~\cite{zeus2009measurement}, and Jefferson Lab~\cite{stepanyan2001observation} have enabled the extraction of quark GPDs. The Jefferson Lab 12 GeV upgrade and the future Electron-Ion Collider (EIC)~\cite{accardi2016electron, AbdulKhalek:2021gbh} are expected to significantly improve our understanding of GPDs, particularly in the gluon sector.

While most studies focus on leading twist GPDs, a complete picture of hadron structure also requires understanding higher twist distributions. Higher twist GPDs, especially twist-3, describe multiparton correlations and include effects such as spin-orbit coupling and transverse force distributions~\cite{belitsky2000twist, kiptily2002twist, penttinen2000deeply}. These correlations are essential for accessing the quark and gluon orbital angular momentum (OAM) and understanding the nucleon's spin decomposition~\cite{ji1997gauge, hatta2012twist, tanaka2019three}. Moreover, twist-3 contributions appear in subleading power corrections in DVCS and may play a role in single-spin asymmetries and transverse structure observables~\cite{burkardt2013transverse}.

In the quark sector, twist-3 GPDs have been investigated in the quark target model~\cite{mukherjee2002off, mukherjee2003helicity}, scalar diquark models~\cite{aslan2020singularities}, and more recently via lattice QCD~\cite{bhattacharya2023chiral, zhang2024twist} and basis light-front quantization (BLFQ)~\cite{zhang2024twist}. However, twist-3 gluon GPDs remain largely unexplored due to their more intricate operator structure, which involves the gluon field strength tensor and its derivatives. These distributions have been discussed in the context of operator identities~\cite{kiptily2002twist}, DVCS amplitude corrections~\cite{belitsky2000twist}, and Wandzura–Wilczek-type relations~\cite{hatta2012twist}, but no comprehensive model-based calculation exists for the full twist-3 gluon GPD sector.

At leading twist, the gluon GPDs include four chiral-even (\( H_g, E_g, \widetilde{H}_g, \widetilde{E}_g \)) and four chiral-odd distributions (\( H_T^g, E_T^g, \widetilde{H}_T^g, \widetilde{E}_T^g \))~\cite{meissner2009generalized}. These have been studied in various phenomenological models and lattice simulations, although the chiral-odd gluon GPDs remain experimentally elusive. Some recent studies have begun probing gluon GTMDs and Wigner distributions~\cite{bacchetta2020gluon, rajabi2022gluonGTMD}, which serve as “mother” distributions to both GPDs and TMDs and are instrumental in exploring spin–orbit and spin–spin correlations.
On the experimental side, accessing gluon GPDs requires processes that are directly sensitive to gluons, such as heavy vector meson production (\( J/\psi \), \( \phi \)) in the small-$x$ regime~\cite{Boussarie:2016qop, Hatta:2018ina}. These processes provide potential access to both leading- and higher-twist gluon GPDs, although isolating twist-3 contributions remains experimentally challenging. Future data from the EIC will be essential in disentangling these effects and constraining gluon contributions to hadron structure.

In this paper, we apply a light-front model~\cite{Sain:2025kup} to investigate the twist-3 chiral-even gluon generalized parton distributions in the proton. The gluon is modeled as the active parton, while the remaining constituents are described as an effective spin-$\tfrac{1}{2}$ spectator system. This setup is motivated by the higher Fock-state decomposition of the nucleon, particularly the \(|qqqg\rangle\) configuration, which naturally incorporates gluon–quark correlations and helicity interference. The light-front wave functions are derived using the soft-wall AdS/QCD model~\cite{vega2009generalized, branz2010softwall}, ensuring confinement dynamics and Regge behavior. We compute the twist-3 gluon GPDs across a wide range of kinematics (\( x, \xi, t \)). We also investigate the potential contribution of twist-3 generalized parton distributions to gluon kinetic OAM and compare them with twist-2 calculations.

The paper is organized as follows. In Section~\ref{Sec:Model}, we present the light-front gluon–spectator model and construct the light-front wave functions. Section~\ref{Sec:Correlator} defines the twist-3 gluon correlators and their overlap representations. In Section~\ref{Sec:Results}, we present and analyze numerical results for twist-3 gluon GPDs and evaluate the gluon kinetic OAM. Section~\ref{Sec:Conclusion} summarizes our findings and outlines possible directions for future work.

\section{Model Description}
\label{Sec:Model}

We consider a light-front two-particle Fock-state model for the proton, in which the gluon is treated as the active parton, while the remaining valence quarks, sea quarks, and additional gluons are collectively described as the spectator system. The light-front wave functions (LFWFs) encode the nonperturbative dynamics of the proton and are expressed in terms of the gluon’s longitudinal momentum fraction \(x\) and its transverse momentum \(\boldsymbol{k}_T\). The two-particle Fock state is denoted by 
\(|\lambda_g,\lambda_X; xP^+, \boldsymbol{k}_T\rangle,\)  
where \(\lambda_g = \pm 1\) and \(\lambda_X = \pm \tfrac{1}{2}\) are the helicities of the gluon and the spectator, respectively. The total helicity of the proton is denoted by \(J_z\). Throughout this work, the notation ``\(\pm\)'' is used to indicate the helicities of the proton, gluon, and spectator.  

To simplify the analysis, we adopt a frame in which the average transverse momentum vanishes, i.e., \(\bar{P}^\perp = 0\). The average momentum is defined as  
\(\bar{P}^\mu = \tfrac{1}{2}\left(P^\mu + P'^\mu\right),\)  
and can be expressed in light-cone coordinates as  
\begin{equation}
    \bar{P}^\mu = p^\mu + \tfrac{1}{2} \tilde{M}^2 n^\mu,
\end{equation}  
where the light-cone vectors \(p^\mu\) and \(n^\mu\) are defined through \(x^\pm = (x^0 \pm x^3)/\sqrt{2}\), and satisfy the normalization condition \(p \cdot n = 1\). The effective squared mass is given by \(\tilde{M}^2 = M^2 - \tfrac{\Delta^2}{4}\). Explicitly, the light-cone basis vectors are chosen as  
\begin{align}
p^\mu &= \frac{1}{\sqrt{2}}(\mathcal{P},0,0,\mathcal{P}), \quad 
n^\mu = \frac{1}{\sqrt{2}}\left(\frac{1}{\mathcal{P}},0,0,-\frac{1}{\mathcal{P}}\right),
\end{align}  
where \(\mathcal{P}\) is a large momentum component along the positive \(z\)-axis, such that \(\bar{P}^+\) represents the dominant light-cone momentum component. The polarization of the target nucleon is described by the spin vector \(S^\mu\), which satisfies the conditions \(S \cdot P = 0\) and \(S^2 = -1\).  

The four-momentum of the active parton (gluon) is  
\(k_1 = \left(x\bar{P}^+, \, \frac{k_1^2 + \boldsymbol{k}_T^2}{x\bar{P}^+}, \, \boldsymbol{k}_T \right),\)  
while the spectator carries  
\(k_2 = \left((1-x)\bar{P}^+, \, k_X^-, \, -\boldsymbol{k}_T \right).\)  

The two-particle Fock-state expansion of the proton with helicity \(J_z = \pm \tfrac{1}{2}\) can be written as \cite{brodsky2001light}  
\begin{align}
    |P(x;\pm)\rangle &= \int \frac{dx \, d^2\boldsymbol{k}_T}{16\pi^3\sqrt{x(1-x)}}
    \Biggl[ \psi^{\pm}_{++}(x,\boldsymbol{k}_T) |+,+;xP^+,\boldsymbol{k}_T\rangle
    + \psi^{\pm}_{+-}(x,\boldsymbol{k}_T) |+,-;xP^+,\boldsymbol{k}_T\rangle \nonumber \\
    &\hspace{2cm}
    + \psi^{\pm}_{-+}(x,\boldsymbol{k}_T) |-,+;xP^+,\boldsymbol{k}_T\rangle
    + \psi^{\pm}_{--}(x,\boldsymbol{k}_T) |-,-;xP^+,\boldsymbol{k}_T\rangle \Biggr],
\end{align}  
where \(\psi^{J_z}_{\lambda_g \lambda_X}\) are the probability amplitudes for the corresponding two-particle states.  

For a proton with \(J_z = +\tfrac{1}{2}\), the LFWFs are  
\begin{align}
    \psi_{++}^{+}(x,\boldsymbol{k}_T) &= -\sqrt{2}\,\frac{-k_1 + i k_2}{x(1-x)}\,\varphi(x,\boldsymbol{k}_T), \nonumber \\
    \psi_{+-}^{+}(x,\boldsymbol{k}_T) &= -\sqrt{2}\left(M - \frac{M_X}{1-x}\right)\varphi(x,\boldsymbol{k}_T), \nonumber \\
    \psi_{-+}^{+}(x,\boldsymbol{k}_T) &= -\sqrt{2}\,\frac{k_1 + i k_2}{x}\,\varphi(x,\boldsymbol{k}_T), \nonumber \\
    \psi_{--}^{+}(x,\boldsymbol{k}_T) &= 0,
    \label{eq:LFWF_Jz+}
\end{align}  
while for \(J_z = -\tfrac{1}{2}\) they are  
\begin{align}
    \psi_{++}^{-}(x,\boldsymbol{k}_T) &= 0, \nonumber \\
    \psi_{+-}^{-}(x,\boldsymbol{k}_T) &= -\sqrt{2}\,\frac{-k_1 + i k_2}{x}\,\varphi(x,\boldsymbol{k}_T), \nonumber \\
    \psi_{-+}^{-}(x,\boldsymbol{k}_T) &= -\sqrt{2}\left(M - \frac{M_X}{1-x}\right)\varphi(x,\boldsymbol{k}_T), \nonumber \\
    \psi_{--}^{-}(x,\boldsymbol{k}_T) &= -\sqrt{2}\,\frac{k_1 + i k_2}{x(1-x)}\,\varphi(x,\boldsymbol{k}_T).
    \label{eq:LFWF_Jz-}
\end{align}  

The common momentum-dependent wave function is modeled as  
\begin{align}
    \varphi(x,\boldsymbol{k}_T) = N_g\,\frac{4\pi}{\kappa} \sqrt{\frac{\log\!\left(\tfrac{1}{1-x}\right)}{x}}\,x^\alpha (1-x)^\beta 
    \exp\!\left[-\frac{\log\!\left(\tfrac{1}{1-x}\right)}{2\kappa^2 x^2}\,\boldsymbol{k}_T^2\right],
    \label{eq:scalar_phi}
\end{align}  
where \(N_g\) is a normalization constant, \(\kappa\) is the AdS/QCD scale parameter, and \(\alpha, \beta\) are phenomenological parameters controlling the small- and large-\(x\) behavior, respectively.  

This parametrization captures the essential features of the gluon--spectator configuration in the proton. The transverse momentum dependence is governed by a logarithmically modified Gaussian distribution inspired by the soft-wall AdS/QCD correspondence \cite{chakrabarti2023gluon,brodsky2015light}, which ensures suppression at large \(\boldsymbol{k}_T\). The small-\(x\) and large-\(x\) dynamics are tuned by the parameters \(\alpha\) and \(\beta\). The model parameters are fixed by fitting NNPDF3.0 data at the scale \(\mu_0 = 2 \ \text{GeV}\) \cite{Sain:2025kup}, and are listed in Table~\ref{tab1}. The spectator mass is set to \(M_X = 0.985^{+0.044}_{-0.045}\,\text{GeV}\), ensuring proton stability \cite{chakrabarti2023gluon}, while the gluon mass is taken to be zero, \(M_g = 0\).  

\begin{table}[h]
\centering
\begin{tabular}{|c|c|c|c|}
\hline
\(N_g\) & \(\kappa \, (\text{GeV})\) & \(\alpha\) & \(\beta\) \\
\hline
\(0.32\) & \(2.62\) & \(-0.530 \pm 0.007\) & \(3.880 \pm 0.223\) \\
\hline
\end{tabular}
\caption{Model parameters fitted to NNPDF3.0 data at \(\mu_0=2 \ \text{GeV}\) \cite{Sain:2025kup}.}
\label{tab1}
\end{table}

\section{Twist-3 GPD of Gluons}
\label{Sec:Correlator}

To investigate the internal spin and momentum structure of the nucleon in terms of gluonic contributions, we study the off-forward matrix elements of bilocal gluon field strength operators connected by a straight Wilson line. The correlators for unpolarized and polarized gluons are defined as \cite{guo2021novel}:
\begin{align}
    F^{g}_{\Lambda^{\prime} \Lambda}(x,\xi,t) &= \int_{-\infty}^{+\infty}\frac{d \lambda}{2 \pi} \, e^{i \lambda x} \, \bigg\langle P',S' \bigg| 2 \, \text{Tr} \left\{ G^{\alpha i}\left(-\tfrac{\lambda n}{2}\right) \mathcal{W}_{-\tfrac{\lambda}{2},\tfrac{\lambda}{2}} G^{\beta}_{\ i}\left(\tfrac{\lambda n}{2}\right) \right\} \bigg|P,S \bigg\rangle, \\
    \tilde{F}^{g}_{\Lambda^{\prime} \Lambda}(x,\xi,t) &= -i \int_{-\infty}^{+\infty} \frac{d \lambda}{2 \pi} \, e^{i \lambda x} \, \bigg\langle P',S' \bigg| 2 \, \text{Tr} \left\{ G^{\alpha i}\left(-\tfrac{\lambda n}{2}\right) \mathcal{W}_{-\tfrac{\lambda}{2},\tfrac{\lambda}{2}} \tilde{G}^{\beta}_{\ i}\left(\tfrac{\lambda n}{2}\right) \right\} \bigg|P,S \bigg\rangle,
\end{align}
where \( G^{\mu\nu} \) denotes the gluon field strength tensor, with its dual defined as \( \tilde{G}^{\mu\nu} = \tfrac{1}{2} \epsilon^{\mu\nu\rho\sigma} G_{\rho\sigma} \). The Wilson line \( \mathcal{W}_{-\lambda/2, \lambda/2} \) ensures gauge invariance of the non-local operator structure. Here, \( P^\mu \) and \( P'^\mu \) denote the four-momenta of the initial and final nucleon states, and the momentum transfer is given by \( \Delta^\mu = P'^\mu - P^\mu \), with invariant momentum transfer \( t = \Delta^2 \).  

The gluon field strength tensor has the standard form
\begin{equation}
    G^{\mu \nu}_a(x) = \partial^{\mu}A_a^{\nu}(x) - \partial^{\nu}A_a^{\mu}(x) + i f_{abc}A_b^{\mu}(x)A_c^{\nu}(x).
\end{equation}

We work in the light-cone gauge, \( A^+=0 \), in which the gauge link becomes trivial. This choice simplifies the calculation since \(G^{+i}(x) = \partial^+ A^i\). The twist-3 parametrization of the gluon correlators can then be expressed in terms of generalized parton distributions (GPDs) as
\begin{align}
    \mathcal{F}^{+\perp}_g&= \bar{P}^{+} \bigg[\frac{\Delta^{\perp}}{M}x G_{g,1}(x,\xi,t) +\Delta^{\perp}\slashed{n} x G_{g,2}(x,\xi,t) +\frac{i \sigma^{\perp\rho}\Delta_{\rho}}{2M}x G_{g,3}(x,\xi,t)\notag\\
    &\hspace{8.5cm}+iM \sigma^{\perp\rho}n_{\rho}x G_{g,4}(x,\xi,t)  \bigg],\label{GPDs1} \\
    \tilde{\mathcal{F}}^{+\perp}_g&= \bar{P}^{+} \bigg[\frac{\Delta^{\perp}\gamma_5}{M}x \tilde{G}_{g,1}(x,\xi,t) +\Delta^{\perp}\slashed{n} \gamma_5 x \tilde{G}_{g,2}(x,\xi,t) +\frac{\gamma_5 \Delta_{\rho}}{2M}x \tilde{G}_{g,3}(x,\xi,t)\notag\\
    &\hspace{8cm}+iM \sigma^{\perp \rho} n_{\rho}\gamma_5 x \tilde{G}_{g,4}(x,\xi,t)  \bigg]\label{GPDs2},
\end{align}
Here, \( G_i(x,\xi,t) \) and \( \tilde{G}_i(x,\xi,t) \) (\(i=1,\dots,4\)) denote the unpolarized and polarized gluon GPDs, respectively.  

Contracting the correlators with nucleon spinors leads to the matrix elements
\begin{align}
    \bar{U}(P',S')F^{\alpha \beta}_gU(P,S)&=\bar{P}^{+}\bar{U}(P',S') \bigg[\frac{\Delta^{\perp}}{M}x G_{g,1}(x,\xi,t) +\Delta^{\perp}\slashed{n} x G_{g,2}(x,\xi,t) +\notag\\
    &\hspace{1cm}\frac{i \sigma^{\perp\rho}\Delta_{\rho}}{2M}x G_{g,3}(x,\xi,t)+iM \sigma^{\perp\rho}n_{\rho}x G_{g,4}(x,\xi,t)  \bigg]U(P,S), \\
    \bar{U}(P',S')\tilde{F}^{\alpha \beta}_g U(P,S)&=\bar{P}^{+}\bar{U}(P',S') \bigg[\frac{\Delta^{\perp}\gamma_5}{M}x \tilde{G}_{g,1}(x,\xi,t) +\Delta^{\perp}\slashed{n} \gamma_5 x \tilde{G}_{g,2}(x,\xi,t) +\notag\\
    &\hspace{1cm}\frac{\gamma_5 \Delta_{\rho}}{2M}x \tilde{G}_{g,3}(x,\xi,t)+iM \sigma^{\perp \rho} n_{\rho}\gamma_5 x \tilde{G}_{g,4}(x,\xi,t)  \bigg] U(P,S).
\end{align}

The Dirac matrices in the light-cone formalism are described in detail in \cite{brodsky1998quantum, harindranath1996introduction}. We now examine the discrete symmetry properties of the GPDs. Under time reversal (\(\mathcal{T}\)) and Hermitian conjugation, the correlators acquire definite transformation properties. In Hilbert space, \(\mathcal{T}\) is an antiunitary operator, acting as
\begin{align}
    \mathcal{T} f &= f^* \mathcal{T}, \\
    \mathcal{T}(f_1+f_2) &= \mathcal{T}f_1 + \mathcal{T}f_2,
\end{align}
with the inner product satisfying
\begin{align}
    \langle\psi_1|\overleftarrow{\mathcal{T}}\overrightarrow{\mathcal{T}}|\psi_2 \rangle = \langle \psi_1 | \psi_2 \rangle^* = \langle \psi_2 | \psi_1 \rangle.
\end{align}

Applying these symmetries to Eqs.~\eqref{GPDs1}--\eqref{GPDs2}, we obtain
\begin{align}
    F(x,\xi,t) &= -F(x,-\xi,t), \label{eq3.10}\\
    \bar{F}(x,\xi,t) &= +\bar{F}(x,-\xi,t),
\end{align}
and under Hermitian conjugation
\begin{align}
    F^*(x,\xi,t) &= -F(x,-\xi,t), \\
    \bar{F}^*(x,\xi,t) &= +\bar{F}(x,-\xi,t). \label{eq3.17}
\end{align}
Here, \(F = \{\tilde{G}_{g,3}, G_{g,1}, G_{g,2}, G_{g,4}\}\) and \(\bar{F} = \{G_{g,3}, \tilde{G}_{g,1}, \tilde{G}_{g,2}, \tilde{G}_{g,4}\}\). From Eqs.~\eqref{eq3.10}--\eqref{eq3.17} we infer that \(xG_{g,3}\), \(x\tilde{G}_{g,1}\), \(x\tilde{G}_{g,2}\), and \(x\tilde{G}_{g,4}\) are even functions of \(\xi\) and real, whereas \(x\tilde{G}_{g,3}\), \(xG_{g,1}\), \(xG_{g,2}\), and \(xG_{g,4}\) are odd functions of \(\xi\) and purely imaginary.  

We proceed within the light-cone quantization framework to express the gluon distributions in terms of light-cone helicity amplitudes. The relevant helicity amplitudes, associated with different nucleon and gluon helicity configurations, are defined as \cite{diehl2003generalized,boffi2007generalized,diehl2001generalized,maji2017leading}:
\begin{align}
    A_{\Lambda'\lambda_g',\Lambda\lambda_g} &= \frac{1}{\bar{P}^+}\int\frac{d\lambda}{2 \pi} e^{i \lambda x} \bigg\langle P',S' \bigg| \epsilon^i(\lambda_g') G^{+i}\left( -\tfrac{\lambda n}{2}\right) G^{+j}\left( +\tfrac{\lambda n}{2}\right)\epsilon^{*j}_{\perp}(\lambda_g) \bigg| P,S\bigg\rangle \Big|_{z^+=0,\boldsymbol{z}_T=0},
\end{align}
where \(\epsilon\) is the transverse gluon polarization vector. Under parity, these amplitudes satisfy
\begin{equation}
    A_{-\Lambda'-\lambda_g',-\Lambda-\lambda_g} = (-1)^{\Lambda'-\lambda_g'-\Lambda+\lambda_g} \left(A_{\Lambda'\lambda_g',\Lambda\lambda_g}\right)^*.
\end{equation}

The twist-3 GPDs can be expressed in terms of the helicity-conserving amplitudes:
\begin{align}
    T_1 &= A_{++,--}+A_{+-,--}+A_{+-,-+}+A_{++,-+}, \label{Helfirst} \\
    T_2 &= A_{--,++}+A_{-+,++}+A_{--,+-}+A_{-+,+-}, \\
    \tilde{T}_1 &= A_{+-,-+}+A_{++,-+}-A_{++,--}-A_{+-,--}, \\
    \tilde{T}_2 &= A_{--,++}+A_{-+,++}-A_{--,+-}-A_{-+,+-}, \\
    \tilde{T}_3 &= A_{++,++}+A_{+-,++}-A_{++,+-}-A_{+-,+-}, \\
    \tilde{T}_4 &= A_{--,-+}+A_{-+,-+}-A_{-+,--}-A_{--,--}. \label{hellast}
\end{align}

Among the eight twist-3 GPDs, only four are even in \(\xi\); thus, in the forward limit \(\xi \to 0\) they survive, while the others vanish. Using the helicity amplitudes \eqref{Helfirst}--\eqref{hellast}, we obtain
\begin{align}
    xG_{g,3}(x,0,\boldsymbol{\Delta}_T^2) &= -\frac{M}{\left(\bar{P}^+\right)^2 \boldsymbol{\Delta}_T^2}\Big\{ (\Delta_1+i \Delta_2)T_1+(\Delta_1-i \Delta_2)T_2\Big\}, \\
    x\tilde{G}_{g,1}(x,0,\boldsymbol{\Delta}_T^2) &= -\frac{M}{2 \bar{P}^+ \boldsymbol{\Delta}_T^3}\Big\{ (\Delta_1+i \Delta_2)\tilde{T}_1+(\Delta_1-i \Delta_2)\tilde{T}_2\Big\}, \\
    x\tilde{G}_{g,2}(x,0,\boldsymbol{\Delta}_T^2) &= \frac{1}{2 x\bar{P}^+}\Big(\tilde{T}_3+\tilde{T}_4\Big), \\
    x\tilde{G}_{g,4}(x,0,\boldsymbol{\Delta}_T^2) &= \frac{ \bar{P}^+}{4 M}\bigg[\frac{\tilde{T}_3}{\boldsymbol{\Delta}_T} + \frac{\tilde{T}_4}{\boldsymbol{\Delta}_T}+ \frac{\boldsymbol{\Delta}_T}{2M (\boldsymbol{k}_T\cdot \boldsymbol{\Delta}_T)-x \bar{P}^+ \boldsymbol{\Delta}_T}(\tilde{T}_1-\tilde{T}_2)\bigg].
\end{align}

The initial and final transverse momenta of the active gluon are
\begin{align}
    \boldsymbol{k}_T^{\prime \prime}= \boldsymbol{k}_T+(1-x)\tfrac{\boldsymbol{\Delta}_T}{2}, \qquad 
    \boldsymbol{k}_T^{\prime}= \boldsymbol{k}_T-(1-x)\tfrac{\boldsymbol{\Delta}_T}{2}.
\end{align}

Finally, employing the light-front wave functions (LFWFs) introduced in Eqs.~\eqref{eq:LFWF_Jz+}--\eqref{eq:LFWF_Jz-}, the twist-3 gluon GPDs at zero skewness are obtained as

\begin{align}
    xG_{g,3}(x,0,-t)&= \frac{4 M}{\boldsymbol{\Delta}_T^2} \int \frac{d^2 \boldsymbol{k}_T}{16 \pi^3} \bigg\{-\frac{(2-x)\boldsymbol{\Delta}_T^2}{x}\left(M-\frac{M_X}{1-x}\right)\notag\\
    &\hspace{2cm}+\left(\frac{1}{x(1-x)}-\frac{1}{x}\right)(\boldsymbol{k}_T \times \boldsymbol{\Delta}_T) \bigg\}
    \varphi^{*}(x,k_T^{\prime \prime})\varphi(x,k_T^{\prime}),
    \label{xG_3}
\end{align}

\begin{align}
    x\tilde{G}_{g,1}(x,0,-t)&= -\frac{4 M}{\boldsymbol{\Delta}_T^3} \int \frac{d^2 \boldsymbol{k}_T}{16 \pi^3} \bigg(M-\frac{M_X}{1-x}\bigg)\frac{1}{1-x}\bigg\{\boldsymbol{k}_T\cdot\boldsymbol{\Delta}_T+\frac{1-x}{2}{\boldsymbol{\Delta}_T^2} \bigg\}\notag\\
    &\hspace{8cm}\varphi^{*}(x,k_T^{\prime \prime})\varphi(x,k_T^{\prime}),
\end{align}

\begin{align}
    x\tilde{G}_{g,2}(x,0,-t)&= \frac{1}{x} \int \frac{d^2 \boldsymbol{k}_T}{16 \pi^3} \bigg\{ \bigg(\frac{1}{x^2(1-x)^2}-\frac{1}{x^2}\bigg)\bigg(\boldsymbol{k}_T^2-\frac{(1-x)^2}{4}\boldsymbol{\Delta}_T^2\bigg) \notag\\
    &\hspace{3cm}+\bigg(M-\frac{M_X}{1-x}\bigg)^2-i\bigg(\frac{1}{x^2(1-x)^2}-\frac{1}{x^2}\bigg)\notag\\
    &\hspace{4.5cm}(1-x)(\boldsymbol{k}_T \times \boldsymbol{\Delta}_T) \bigg\}
    \varphi^{*}(x,k_T^{\prime \prime})\varphi(x,k_T^{\prime}),
\end{align}

\begin{align}
    x \tilde{G}_{g,4}(x, 0, t) &=
    \frac{x}{2 M^2 \boldsymbol{\Delta}_T}
    \int \frac{d^2 \boldsymbol{k}_T}{16\pi^3} \, \Bigg\{ 
    \left[\frac{2}{x^2 (1 - x)^2}
    \left( \boldsymbol{k}_T^2 - \frac{(1 - x)^2}{4} \boldsymbol{\Delta}_T^2 
    \right)\right.\notag \\
    &\hspace{1cm}\left.-\left( M - \frac{M_X}{1 - x} \right)^2
    \right] +  \left[ \bigg( M - \frac{M_X}{1 - x} \right)^2-\left( \frac{1}{(1 - x)^2} \right.\notag \\
    &\hspace{1.5cm} + \frac{1}{x^2 (1 - x)^2}
    \bigg)\left( \boldsymbol{k}_T^2 - \frac{(1 - x)^2}{4} \boldsymbol{\Delta}_T^2 \right)\bigg] \Bigg\} 
    \varphi^*(x, \boldsymbol{k}_T'') \varphi(x, \boldsymbol{k}_T').
\end{align}

\section{Results and Discussions}
\label{Sec:Results}

In this section we present our numerical findings of twist-3 gluon GPDs \(xG_i\) and \(x \tilde{G}_i\) using the light front model. We also present the twist-3 IPDPDFs and the gravitational form factors associated with twist-3 GPDs. In the last section the twist-3 PDF \(\Delta G_T\) findings are presented.

\subsection{GPDs}

At zero skewness, the symmetry properties of the system ensure that only four generalized parton distributions (GPDs) remain non-vanishing. In this section, we present results for the GPDs \(xG_{g,3}(x,\boldsymbol{\Delta}_T^2)\), \(x\tilde{G}_{g,1}(x,\boldsymbol{\Delta}_T^2)\), \(x\tilde{G}_{g,2}(x,\boldsymbol{\Delta}_T^2)\), and \(x\tilde{G}_{g,4}(x,\boldsymbol{\Delta}_T^2)\) within the kinematic ranges \(x=0.05\text{--}0.6\) and \(\boldsymbol{\Delta}_T^2=0.01\text{--}2\,\mathrm{GeV}^2\).

\begin{figure*}[htbp]
    \centering
    \begin{subfigure}[b]{0.45\textwidth}
        \centering
        \includegraphics[width=\textwidth]{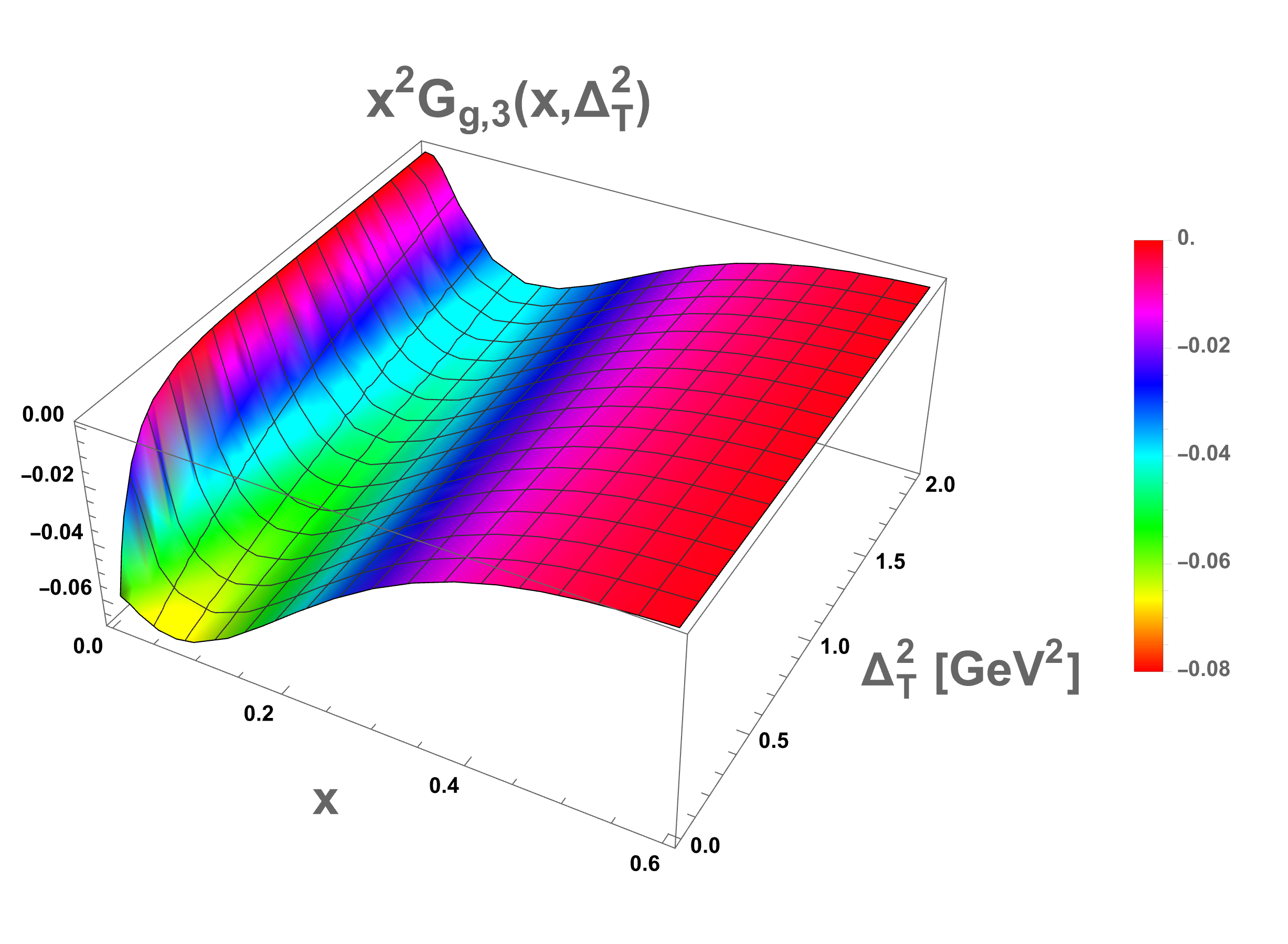} 
        \caption{}
    \end{subfigure}
    \hspace{1cm}
    \begin{subfigure}[b]{0.45\textwidth}
        \centering
        \includegraphics[width=\textwidth]{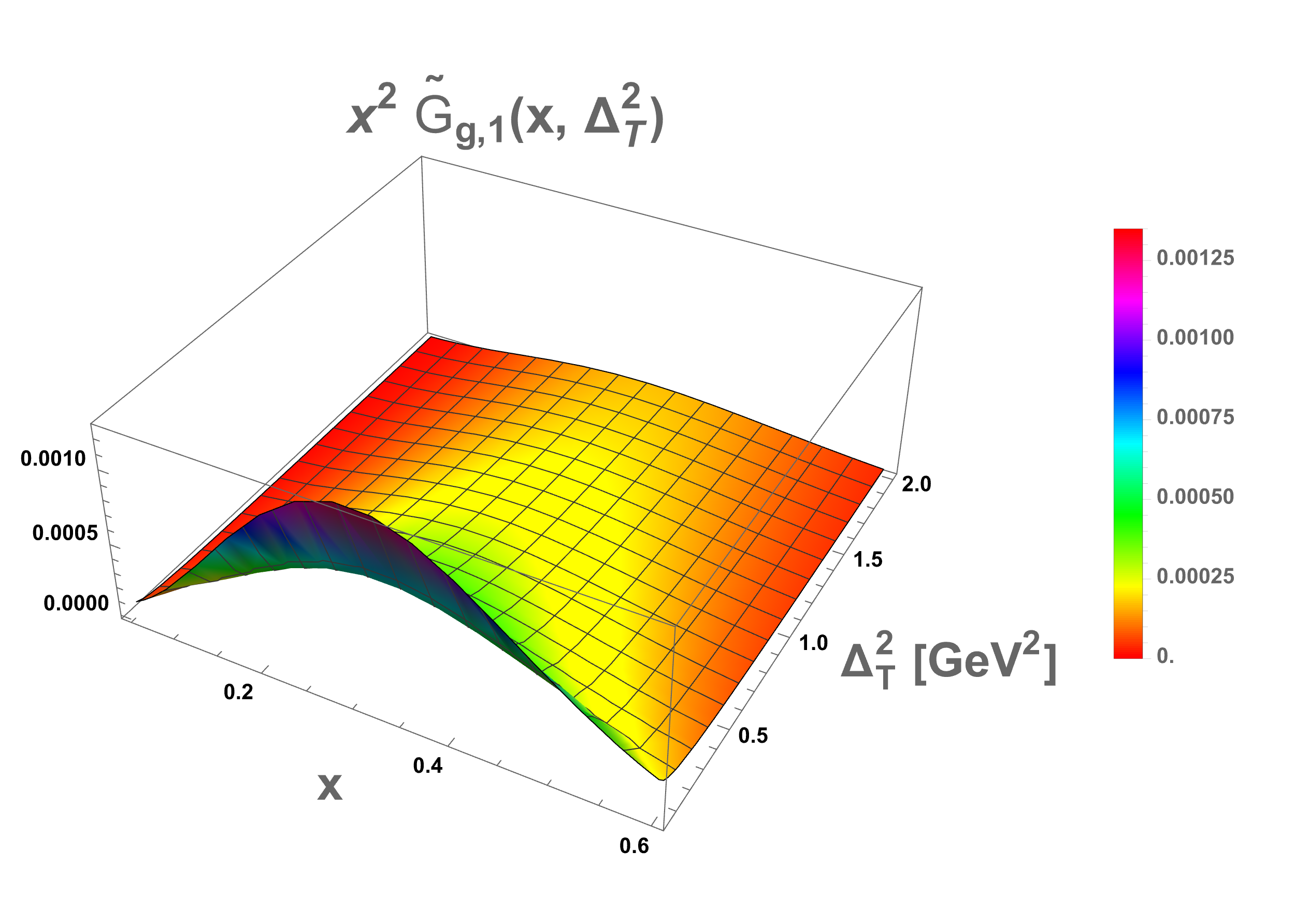} 
        \caption{}
    \end{subfigure}

    \vspace{1em}

    \begin{subfigure}[b]{0.45\textwidth}
        \centering
        \includegraphics[width=\textwidth]{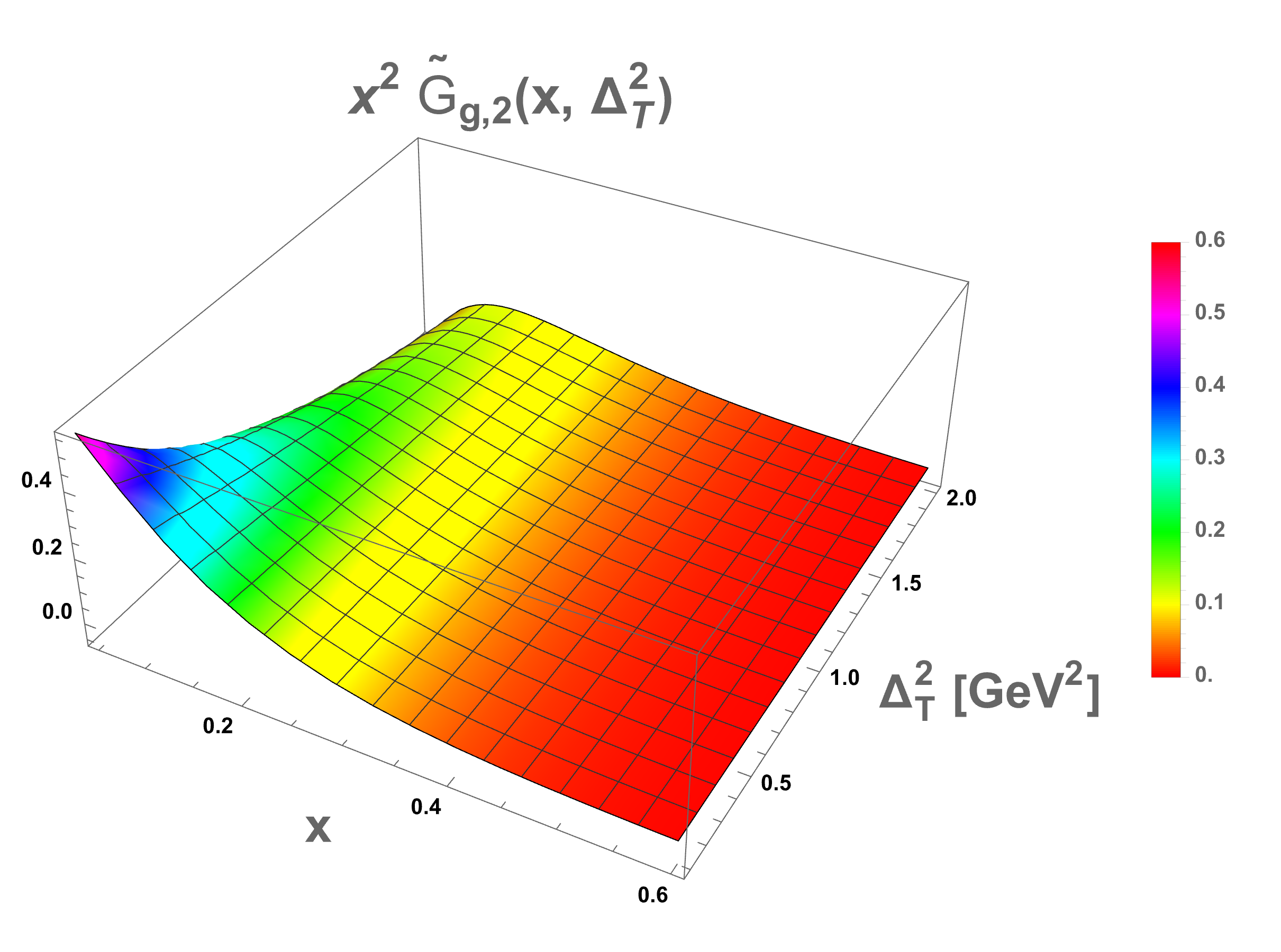} 
        \caption{}
    \end{subfigure}
    \hspace{1cm}
    \begin{subfigure}[b]{0.45\textwidth}
        \centering
        \includegraphics[width=\textwidth]{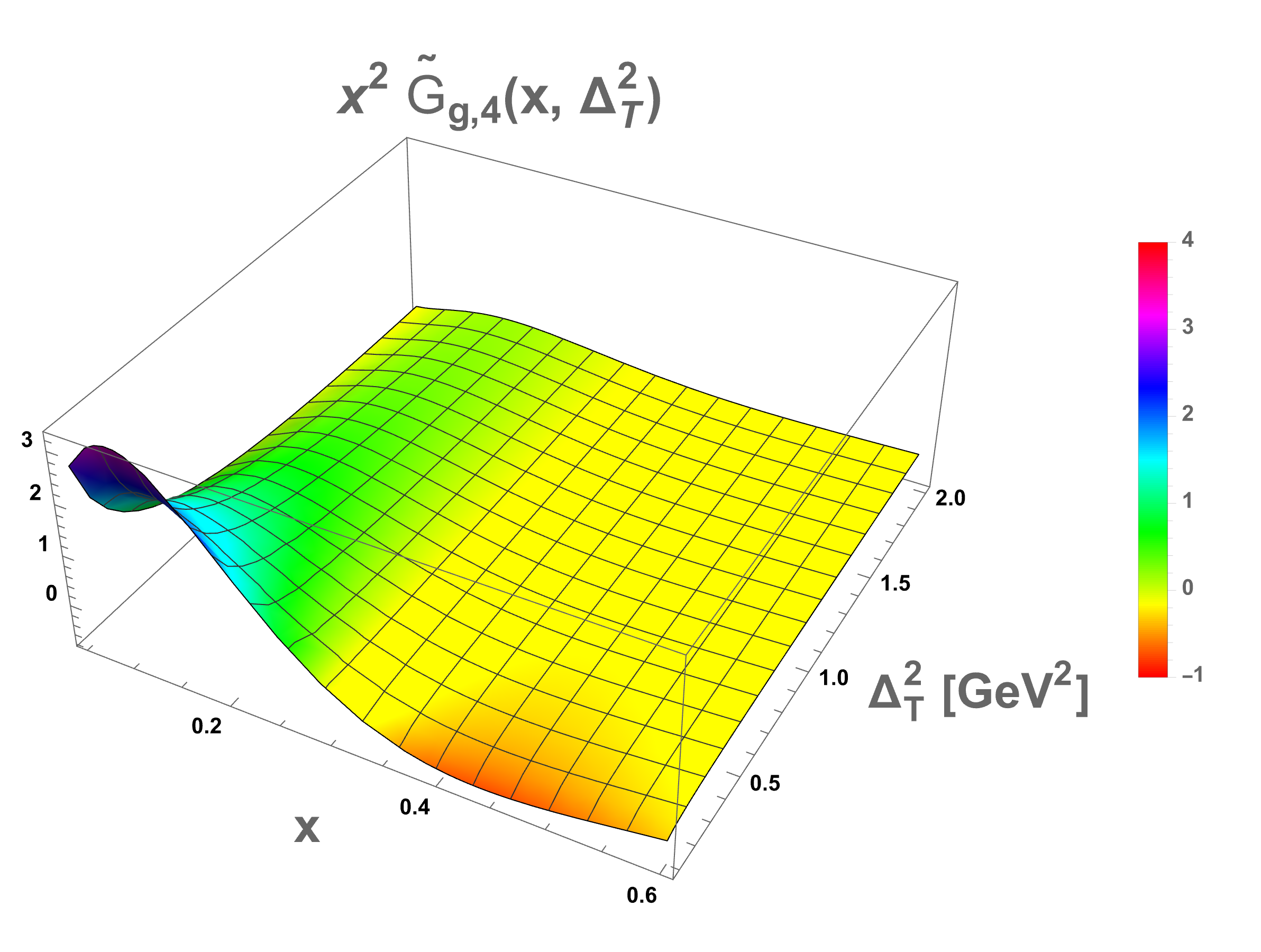} 
        \caption{}
    \end{subfigure}
    \caption{The twist-3 GPDs \(x {G}_{g,3}(x,\boldsymbol{\Delta}_T^2)\),\(x{\tilde{G}}_{g,1}(x,\boldsymbol{\Delta}_T^2)\),\(x{\tilde{G}}_{g,2}(x,\boldsymbol{\Delta}_T^2)\) and \(x{\tilde{G}}_{g,4}(x,\boldsymbol{\Delta}_T^2)\) are plotted with respect to \(x\) and \(\boldsymbol{\Delta}_T^2[\text{GeV}^2]\) in the kinematic range \(x\in[0.05,0.6]\) and \(\boldsymbol{\Delta}_T^2\in[0.01,2] \ \text{GeV}^2\).}
    \label{Figure1}
\end{figure*}

Figure~\ref{Figure1} shows the model predictions for the twist-3 gluon GPDs 
$x^2\tilde{G}_{g,1}(x,\Delta_T^2)$, $x^2\tilde{G}_{g,2}(x,\Delta_T^2)$, $x^2\tilde{G}_{g,4}(x,\Delta_T^2)$, 
and $x^2 G_{g,3}(x,\Delta_T^2)$ as functions of the gluon momentum fraction $x$ and the squared transverse 
momentum transfer $\Delta_T^2$. All four GPDs exhibit a strong enhancement at small $x$ and low $\Delta_T^2$, 
followed by a monotonic suppression with increasing $\Delta_T^2$, reflecting the expected form-factor--like 
fall-off in impact-parameter space. The magnitude ordering 
$\tilde{G}_{g,4} \gg \tilde{G}_{g,2} \gg G_{g,3}\gg \tilde{G}_{g,1} $ highlights the relative contributions 
of the different twist-3 gluonic structures. While $\tilde{G}_{g,1}$ is positive but numerically suppressed, 
$\tilde{G}_{g,2}$ remains sizable and positive over a wide $x$ range, and $\tilde{G}_{g,4}$ attains the largest 
magnitude with a sign change at intermediate $x$, the distribution $G_{g,3}$ is negative throughout the entire 
kinematic domain. The latter is of particular interest since its $x$- and $\Delta_T^2$-dependence directly enters 
the twist-3 sector of the gluon gravitational form factor, which is related to the mechanical properties of the 
nucleon such as the distribution of pressure and shear forces.

\begin{figure*}[htbp]
    \centering
    \begin{subfigure}[b]{0.45\textwidth}
        \centering
        \includegraphics[width=\textwidth]{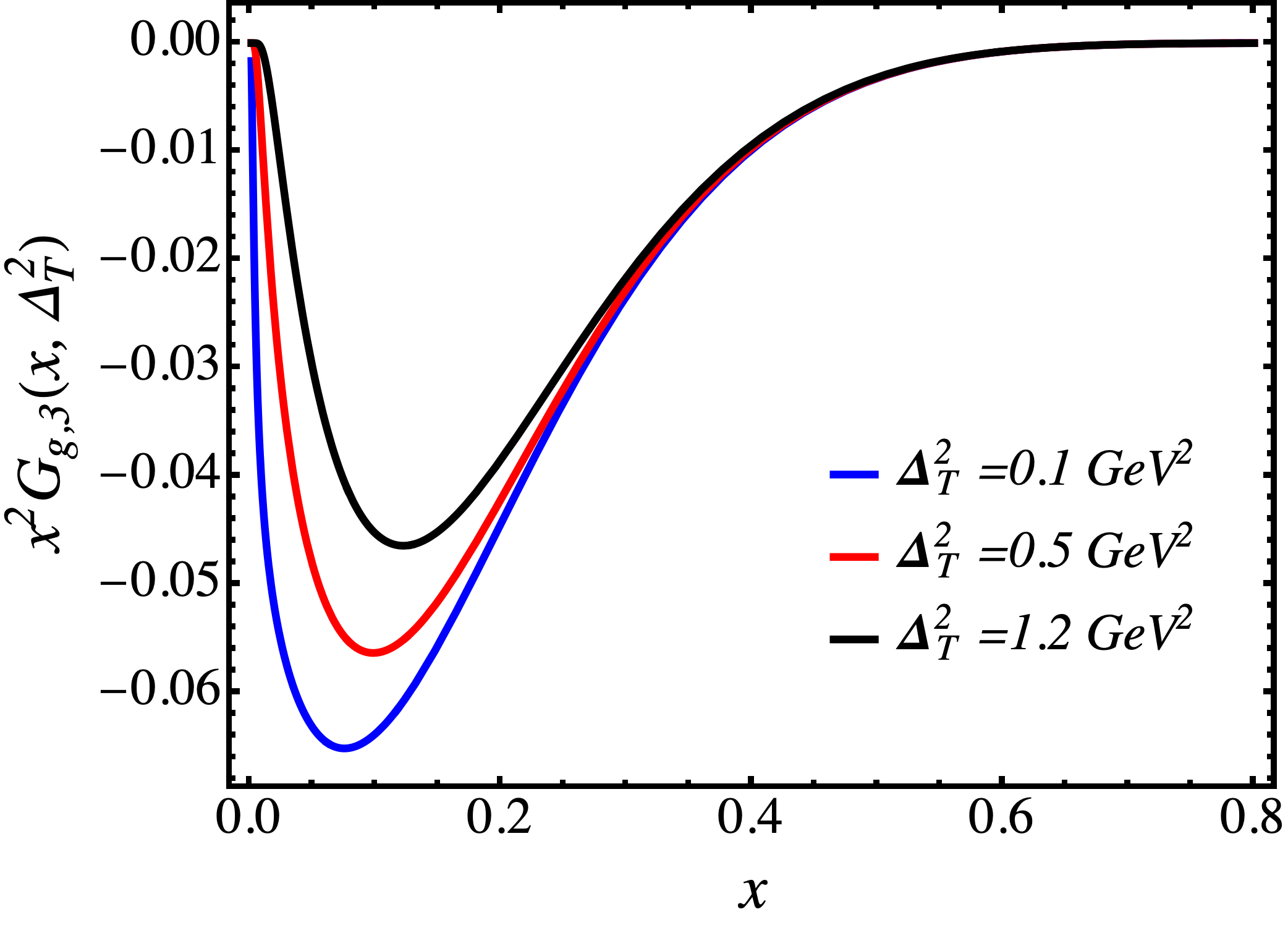} 
        \caption{}
    \end{subfigure}
    \hspace{1cm}
    \begin{subfigure}[b]{0.45\textwidth}
        \centering
        \includegraphics[width=\textwidth]{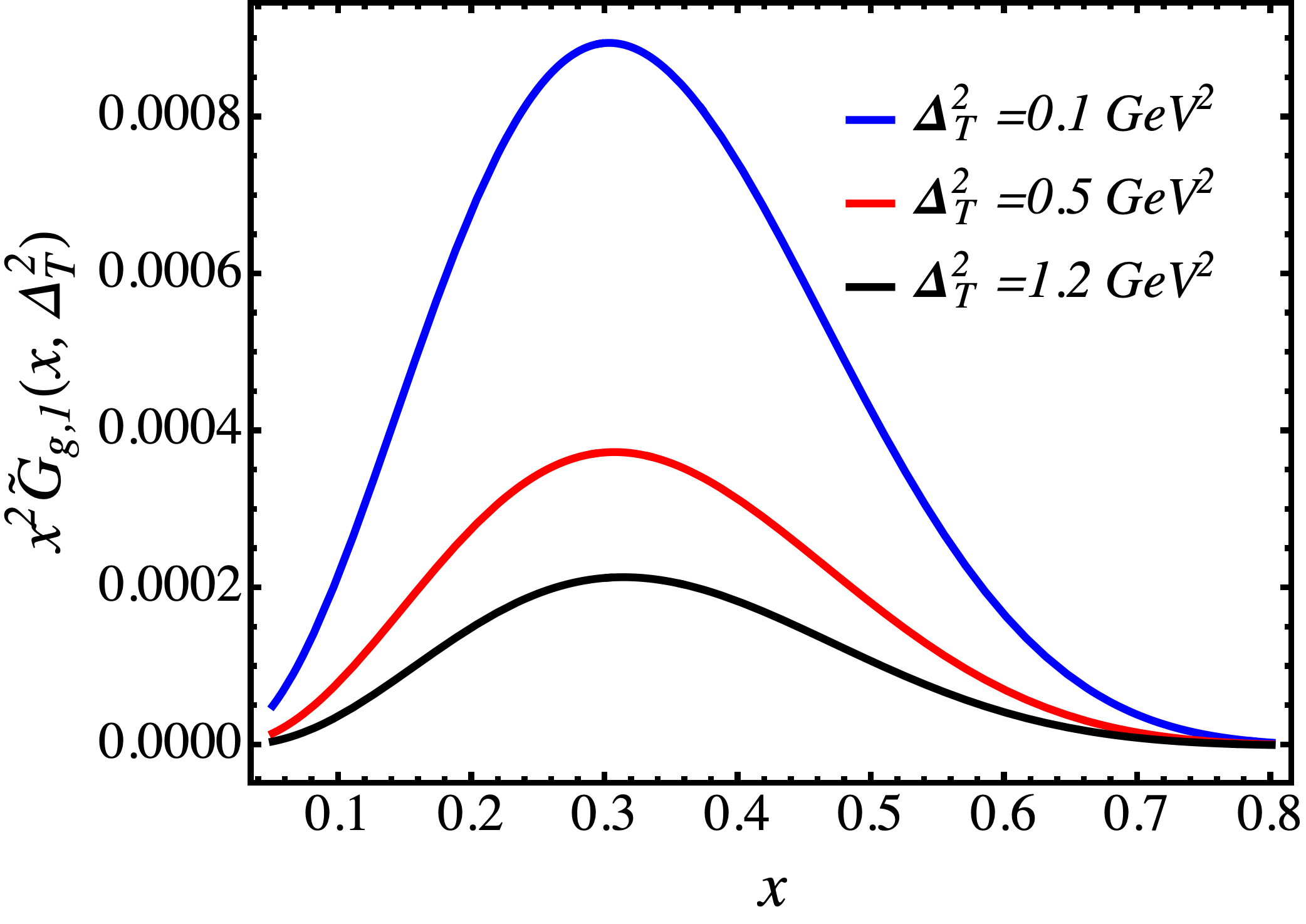} 
        \caption{}
    \end{subfigure}

    \vspace{1em}

    \begin{subfigure}[b]{0.45\textwidth}
        \centering
        \includegraphics[width=\textwidth]{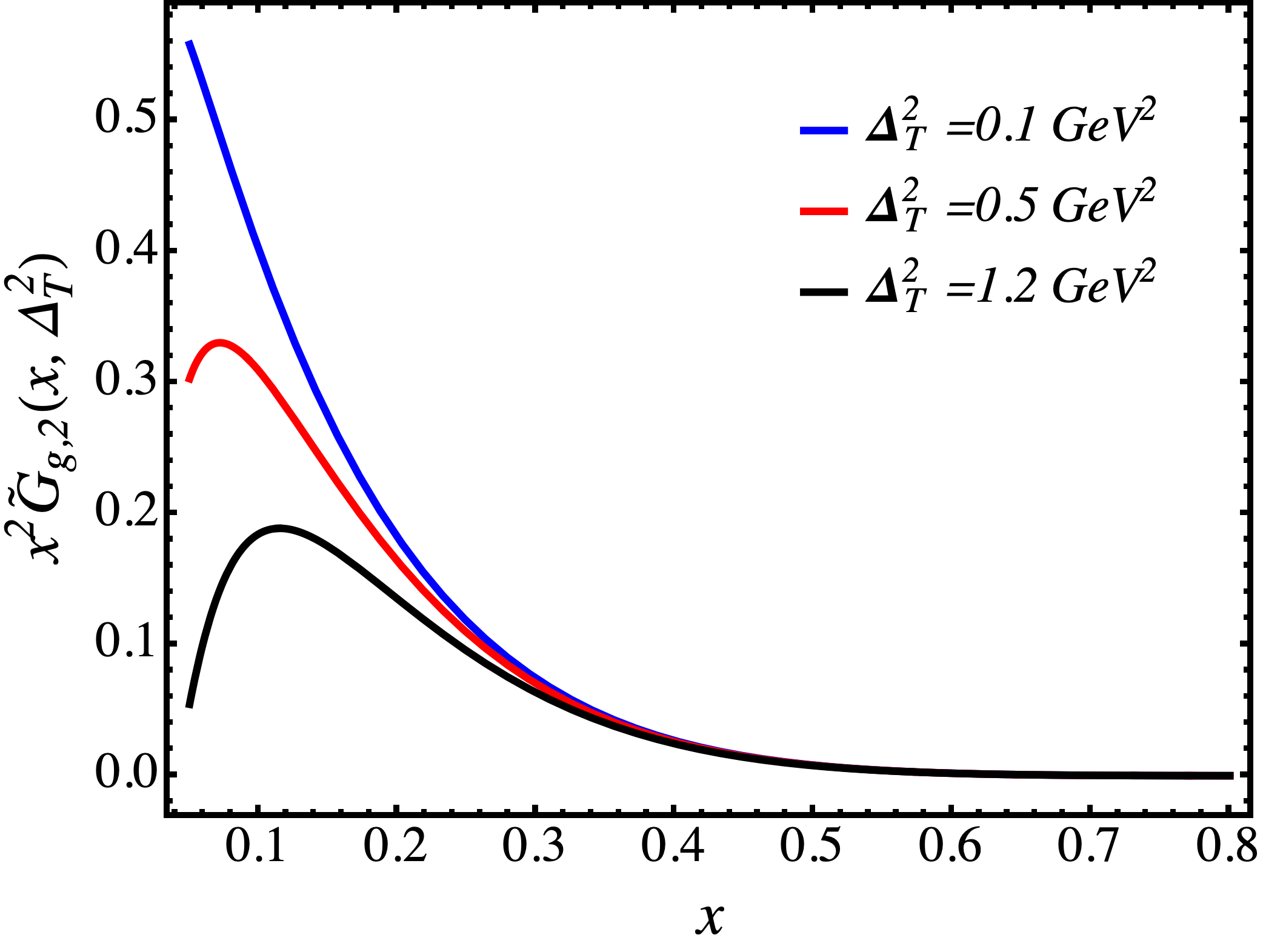} 
        \caption{}
    \end{subfigure}
    \hspace{1cm}
    \begin{subfigure}[b]{0.45\textwidth}
        \centering
        \includegraphics[width=\textwidth]{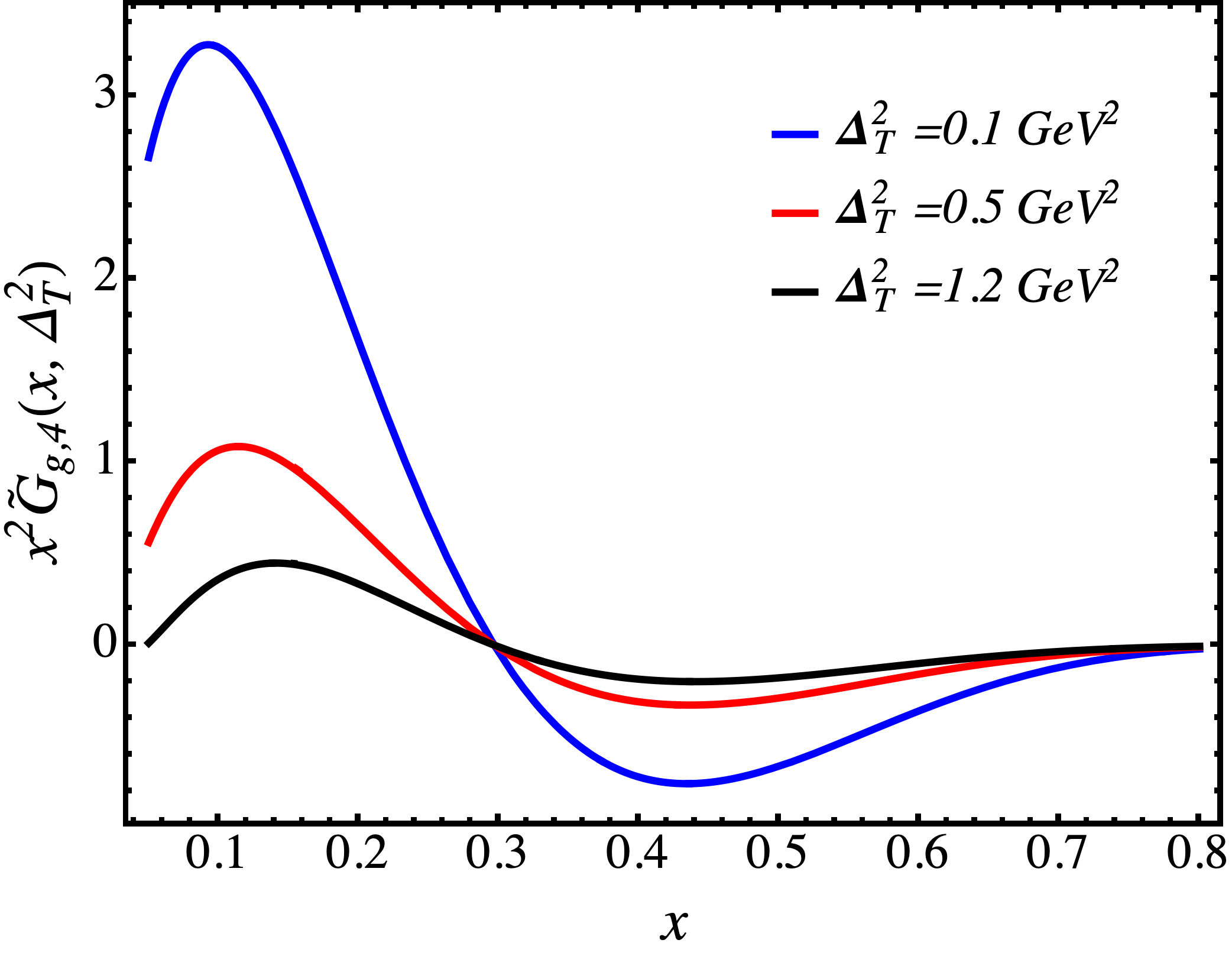} 
        \caption{}
    \end{subfigure}
    \caption{The twist-3 GPDs \(x {G}_{g,3}(x,\boldsymbol{\Delta}_T^2)\),\(x{\tilde{G}}_{g,1}(x,\boldsymbol{\Delta}_T^2)\),\(x{\tilde{G}}_{g,2}(x,\boldsymbol{\Delta}_T^2)\) and \(x{\tilde{G}}_{g,4}(x,\boldsymbol{\Delta}_T^2)\) are plotted with respect to \(x\) in the kinematic range \(x\in[0.05,0.8]\) at fixed of \(\boldsymbol{\Delta}_T^2=0.1(\text{blue}) \ [\text{GeV}^2],0.5(\text{red})\ [\text{GeV}^2]\) and \(\boldsymbol{\Delta}_T^2=1.2 \ [\text{GeV}^2](\text{black})\).}
    \label{Figure2}
\end{figure*}

Figure~\ref{Figure2} show the $x$-dependence of the scaled twist-3 gluon GPDs 
$x^2\tilde{G}_{g,2}$, $x^2\tilde{G}_{g,1}$, $x^2 G_{g,3}$, and $x^2\tilde{G}_{g,4}$ for 
$\Delta_T^2 = 0.1,\ 0.5,$ and $1.2\ \mathrm{GeV}^2$. In all cases, increasing $\Delta_T^2$ suppresses the 
magnitude, consistent with a form-factor--like fall-off reflecting the transverse localization of gluons in the nucleon.  

At small $x$, $\tilde{G}_{g,2}$ shows a strong enhancement, indicative of gluon density growth in the Regge regime, 
while all distributions vanish toward $x \to 1$ due to phase-space suppression. $\tilde{G}_{g,1}$ exhibits a bell-shaped peak at intermediate $x$, whose height and width shrink with $\Delta_T^2$.  
$G_{g,3}$ is negative-definite, with its magnitude largest at small $x$; its connection to twist-3 gluon gravitational form factors makes this suppression relevant for the study of pressure and shear distributions.  
$\tilde{G}_{g,4}$ changes sign near $x \approx 0.3$, with both positive and negative lobes reduced at higher $\Delta_T^2$ but a stable zero-crossing point.  

While the detailed shapes are model dependent, the observed sign structures, small-$x$ enhancements, and  $\Delta_T^2$ suppression patterns are robust consequences of twist-3 gluon spin--orbit and spin--spin correlations.  
The use of 2D $(x,\Delta_T^2)$ slices of the full 3D GPDs (in $(x,\xi,\Delta_T^2)$ space) provides direct insight into the interplay between longitudinal momentum and transverse spatial structure, enabling a clearer interpretation of the dynamical mechanisms governing subleading-twist gluon dynamics.

\subsection{Parton Distributions in Impact Parameter Space}

The spatial structure of partons within the nucleon can be explored by studying generalized parton distributions (GPDs) in the impact parameter space also known as impact parameter dependent parton distributions (IPDPDFs). This formalism allows for a simultaneous description of both the longitudinal momentum fraction and the transverse spatial distribution of partons~\cite{Burkardt:2000za,Diehl:2002he}.

The distributions in impact parameter space are obtained by performing a 2-D Fourier transform of the GPDs with respect to the transverse momentum transfer \(\boldsymbol{\Delta}_T\). For gluons, the impact parameter dependent distributions are defined as:
\begin{align}
    x\,\mathcal{G}_{g,i}\left(x, \boldsymbol{b}_T\right) &= \int \frac{d^2 \boldsymbol{\Delta}_T}{(2\pi)^2}\, e^{-i \boldsymbol{\Delta}_T \cdot \boldsymbol{b}_T}\, xG_{g,i}\left(x, \boldsymbol{\Delta}_T^2\right), \\
    x\,\tilde{\mathcal{G}}_{g,i}\left(x, \boldsymbol{b}_T\right) &= \int \frac{d^2 \boldsymbol{\Delta}_T}{(2\pi)^2}\, e^{-i \boldsymbol{\Delta}_T \cdot \boldsymbol{b}_T}\, x \tilde{G}_{g,i}\left(x, \boldsymbol{\Delta}_T^2\right),
\end{align}

where $\boldsymbol{b}_T$ denotes the transverse position (impact parameter) of the parton relative to the nucleon's transverse center of momentum. These impact parameter-dependent distributions provide a three-dimensional picture of the partonic structure of the nucleon and are particularly useful in elucidating correlations between the momentum and spatial distributions of gluons.

\begin{figure*}[htbp]
    \centering
    \begin{subfigure}[b]{0.45\textwidth}
        \centering
        \includegraphics[width=\textwidth]{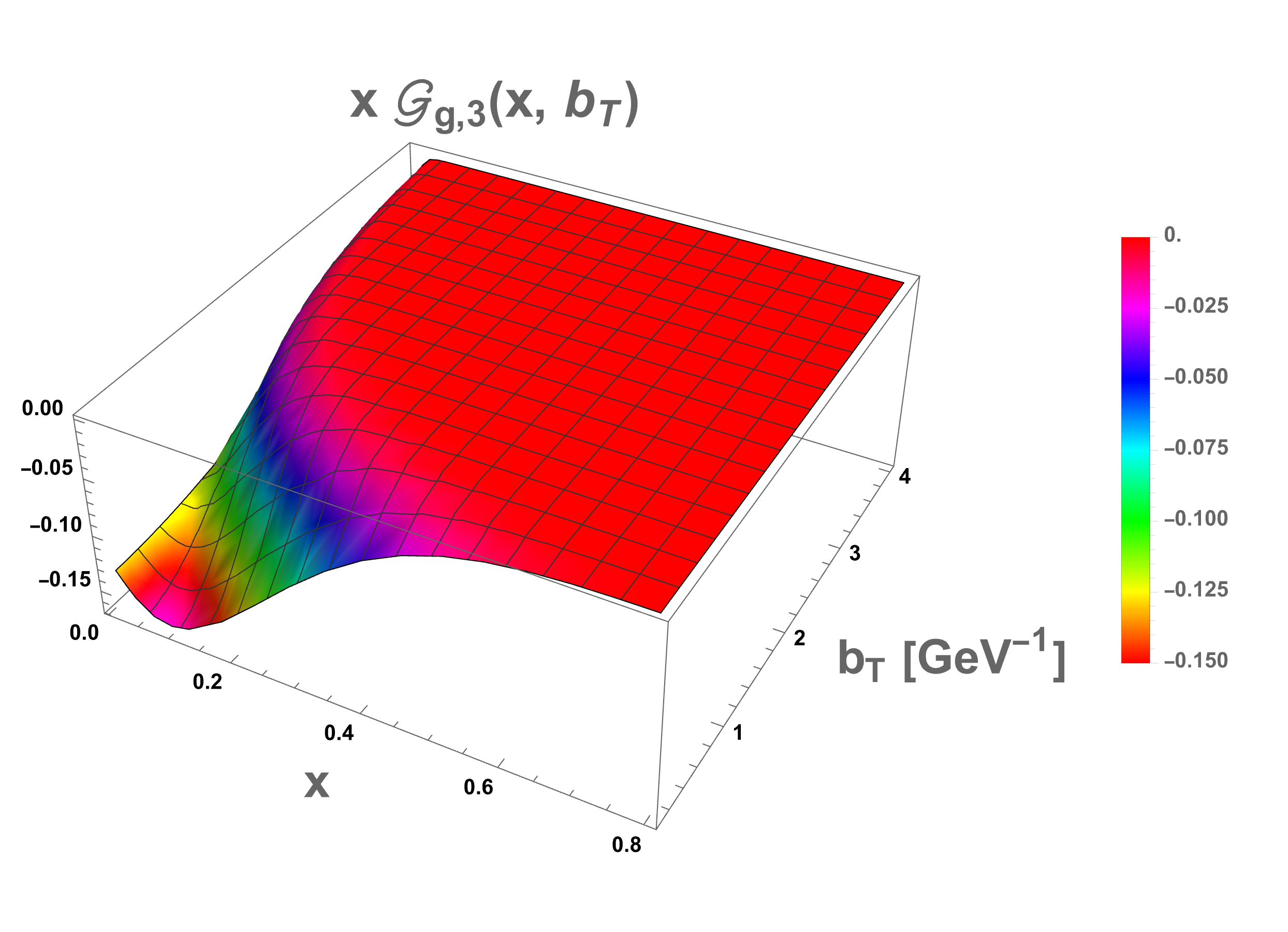} 
        \caption{}
    \end{subfigure}
    \hspace{1cm}
    \begin{subfigure}[b]{0.45\textwidth}
        \centering
        \includegraphics[width=\textwidth]{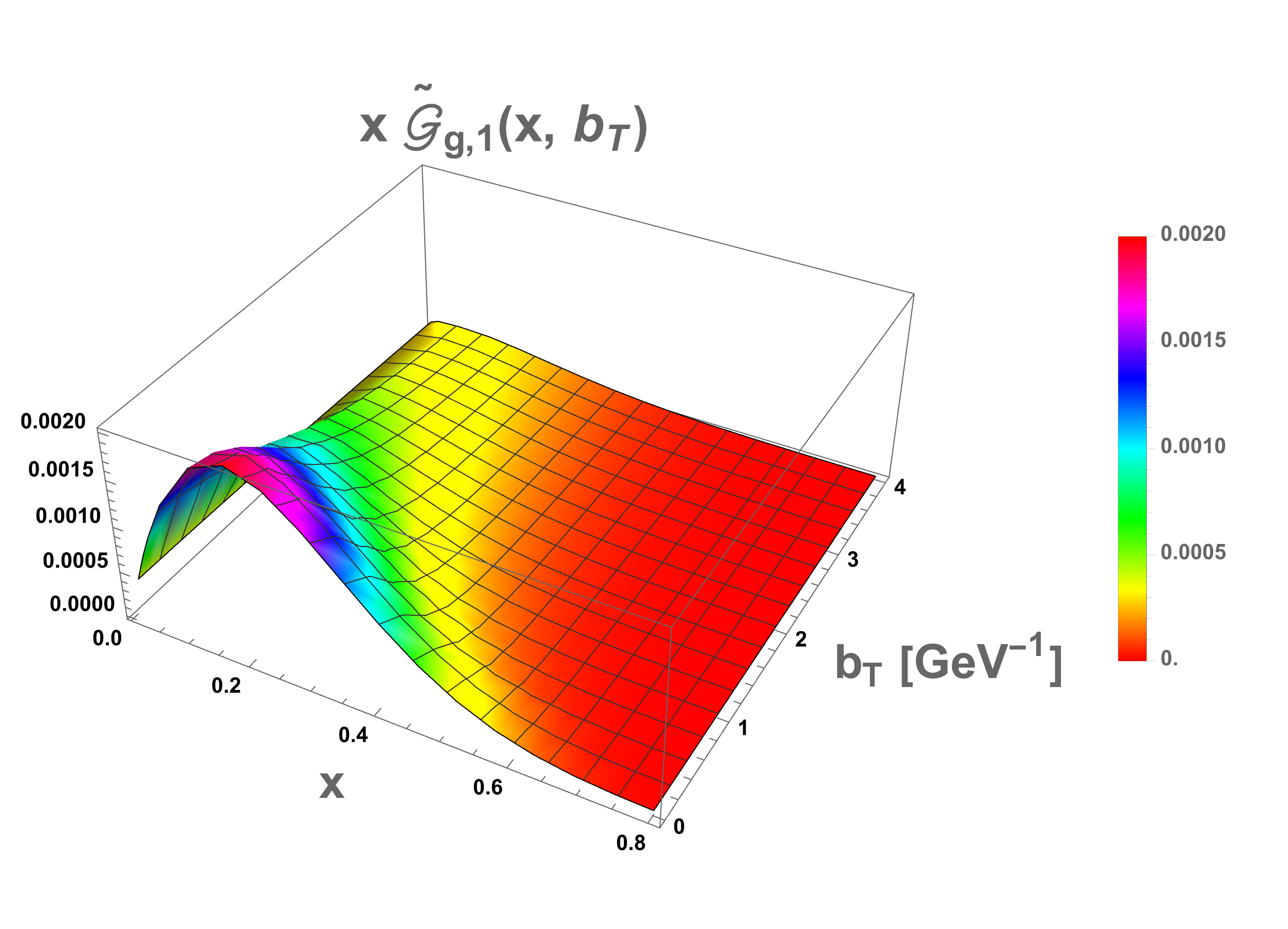} 
        \caption{}
    \end{subfigure}

    \vspace{1em}

    \begin{subfigure}[b]{0.45\textwidth}
        \centering
        \includegraphics[width=\textwidth]{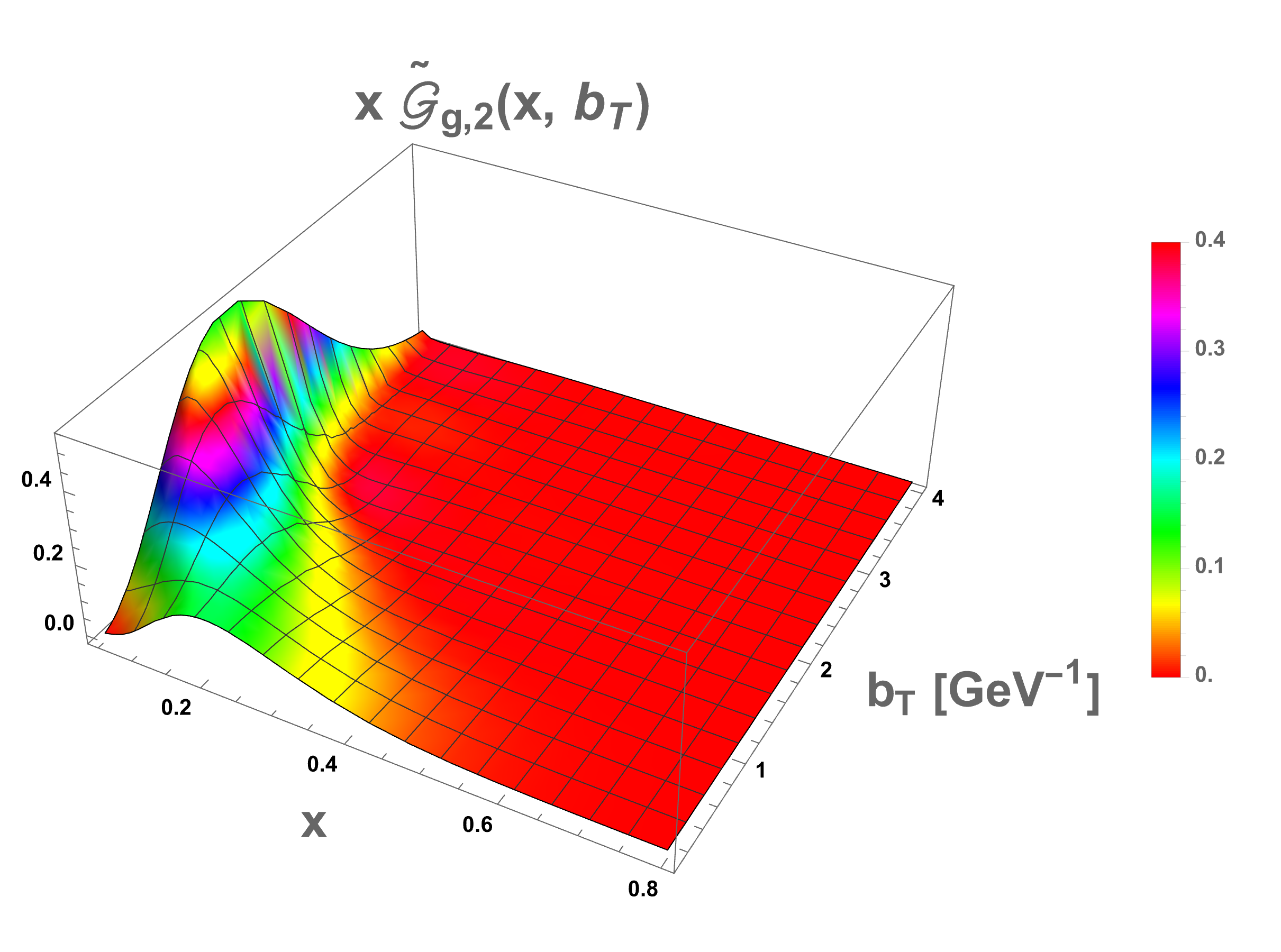} 
        \caption{}
    \end{subfigure}
    \hspace{1cm}
    \begin{subfigure}[b]{0.45\textwidth}
        \centering
        \includegraphics[width=\textwidth]{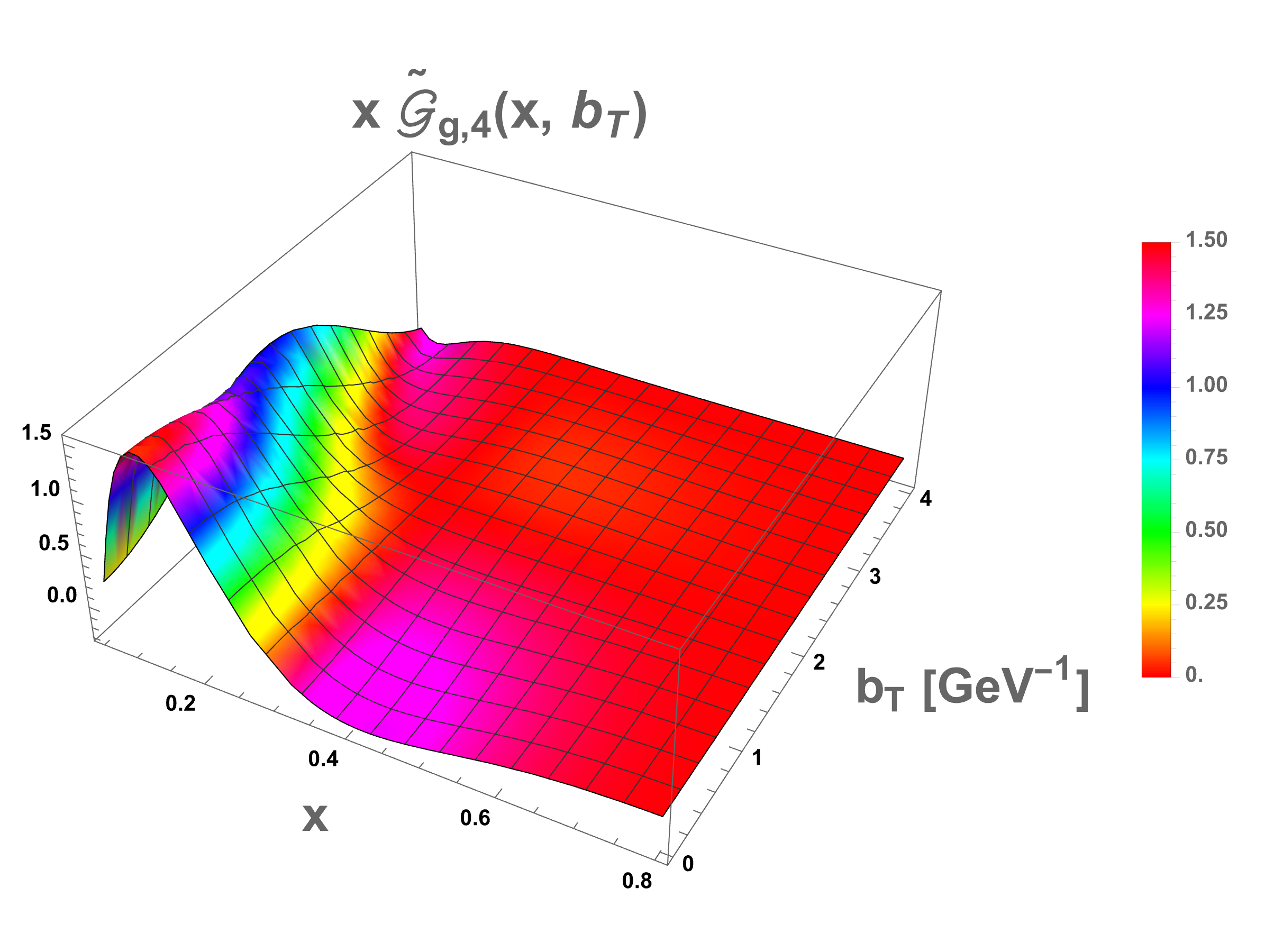} 
        \caption{}
    \end{subfigure}
    \caption{The twist-3 IPDPDFs \(x \mathcal{G}_{g,3}(x,b_T)\),\(x\mathcal{\tilde{G}}_{g,1}(x,b_T)\),\(x\mathcal{\tilde{G}}_{g,2}(x,b_T)\) and \(x\mathcal{\tilde{G}}_{g,4}(x,b_T)\) are plotted with respect to \(x\) and \(b_T [\text{GeV}^{-1}]\) in the kinematic range \(x\in[0.05,0.8]\) and \(b_T\in[0.01,4]\).}
    \label{Figure3}
\end{figure*}

The 3D distributions, correlating the longitudinal momentum fraction $x$ and the transverse spatial coordinate $b_T$, are shown in figure~\ref{Figure3}, offering a tomographic view of the proton's gluon content. $x\mathcal{G}_{g,3}(x, b_T)$, shown in Fig.~\ref{Figure3}(a), is entirely negative, reflecting specific gluon spin-orbit correlations. The peak in magnitude occurs at low $x$ and $b_T=0$, but the distribution is broader in both $x$ and $b_T$, implying a more diffuse spatial arrangement. Figure~\ref{Figure3}(b) displays the distribution for $x\mathcal{\tilde{G}}_{g,1}(x, b_T)$, which is positive definite. It peaks at $x \approx 0.1$ and $b_T = 0$, falling off rapidly in $b_T$, which indicates a central concentration of gluons. The distribution's prominence at low $x$ aligns with the expected behavior of gluon densities. The distribution for $x\mathcal{\tilde{G}}_{g,2}(x, b_T)$ in Fig.~\ref{Figure3}(c) is also positive but shows a sharper and larger peak at very low $x \approx 0.05$. Its narrower $b_T$ profile suggests an even stronger central localization compared to $\mathcal{\tilde{G}}_{g,1}$. Finally, Fig.~\ref{Figure3}(d) shows $x\mathcal{\tilde{G}}_{g,4}(x, b_T)$, which is positive and has the largest magnitude of the four GPDs. The distribution is highly localized, with a pronounced peak at $x \approx 0.05$ and $b_T \approx 0$, underscoring its significant role in the small-$x$ dynamics of the proton.

\begin{figure*}[htbp]
    \centering
    \begin{subfigure}[b]{0.45\textwidth}
        \centering
        \includegraphics[width=\textwidth]{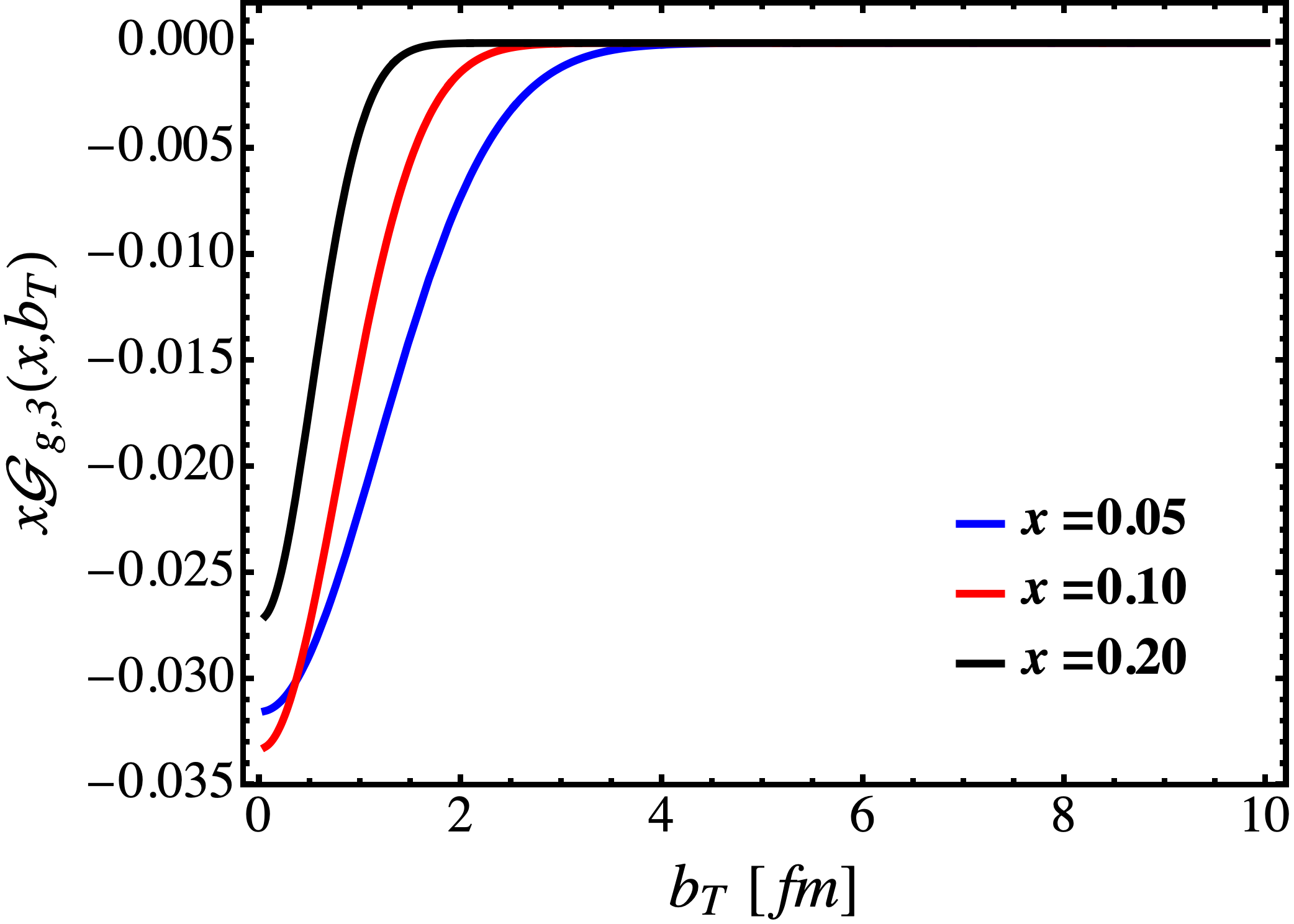} 
        \caption{}
    \end{subfigure}
    \hspace{1cm}
    \begin{subfigure}[b]{0.45\textwidth}
        \centering
        \includegraphics[width=\textwidth]{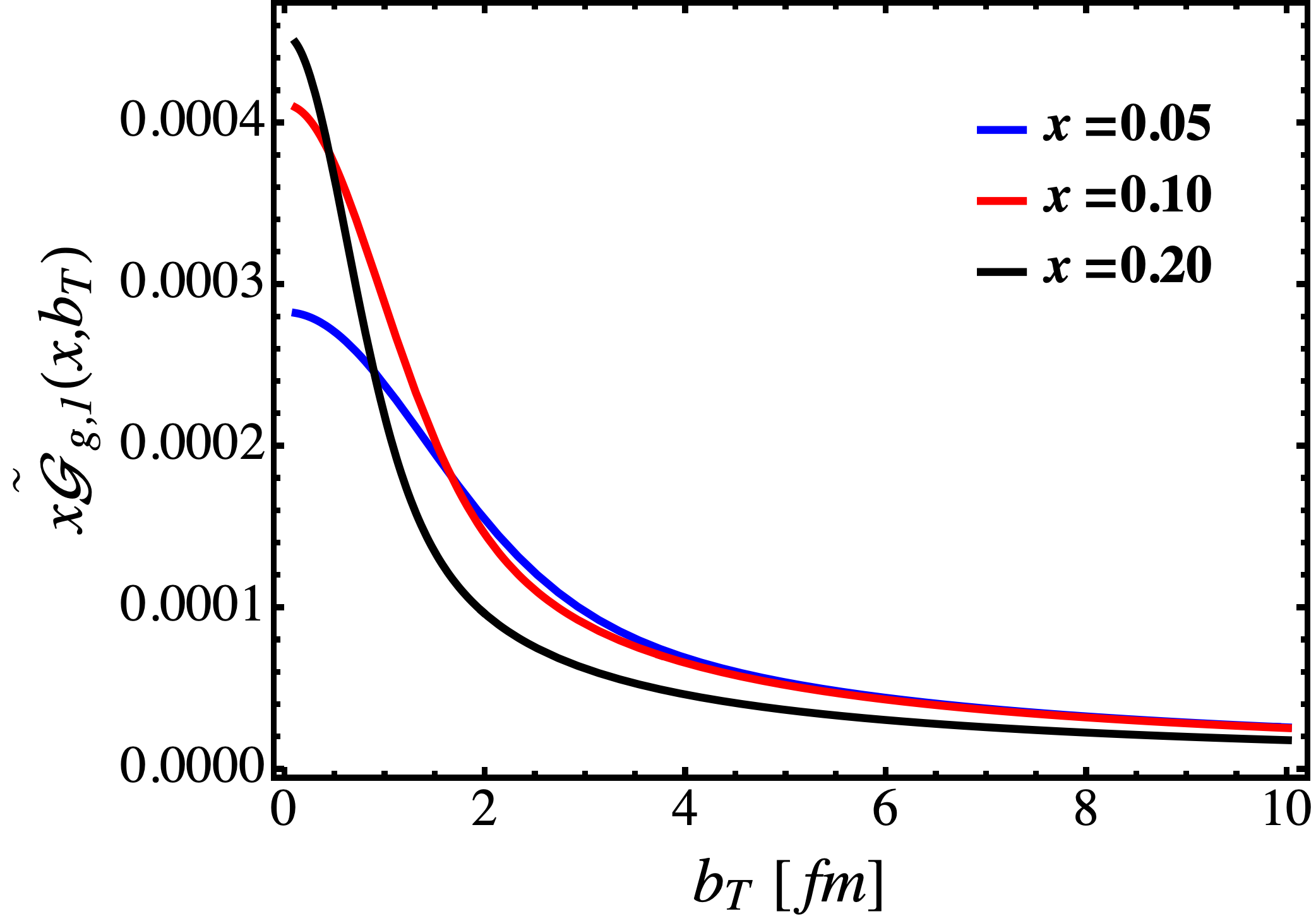} 
        \caption{}
    \end{subfigure}

    \vspace{1em}

    \begin{subfigure}[b]{0.45\textwidth}
        \centering
        \includegraphics[width=\textwidth]{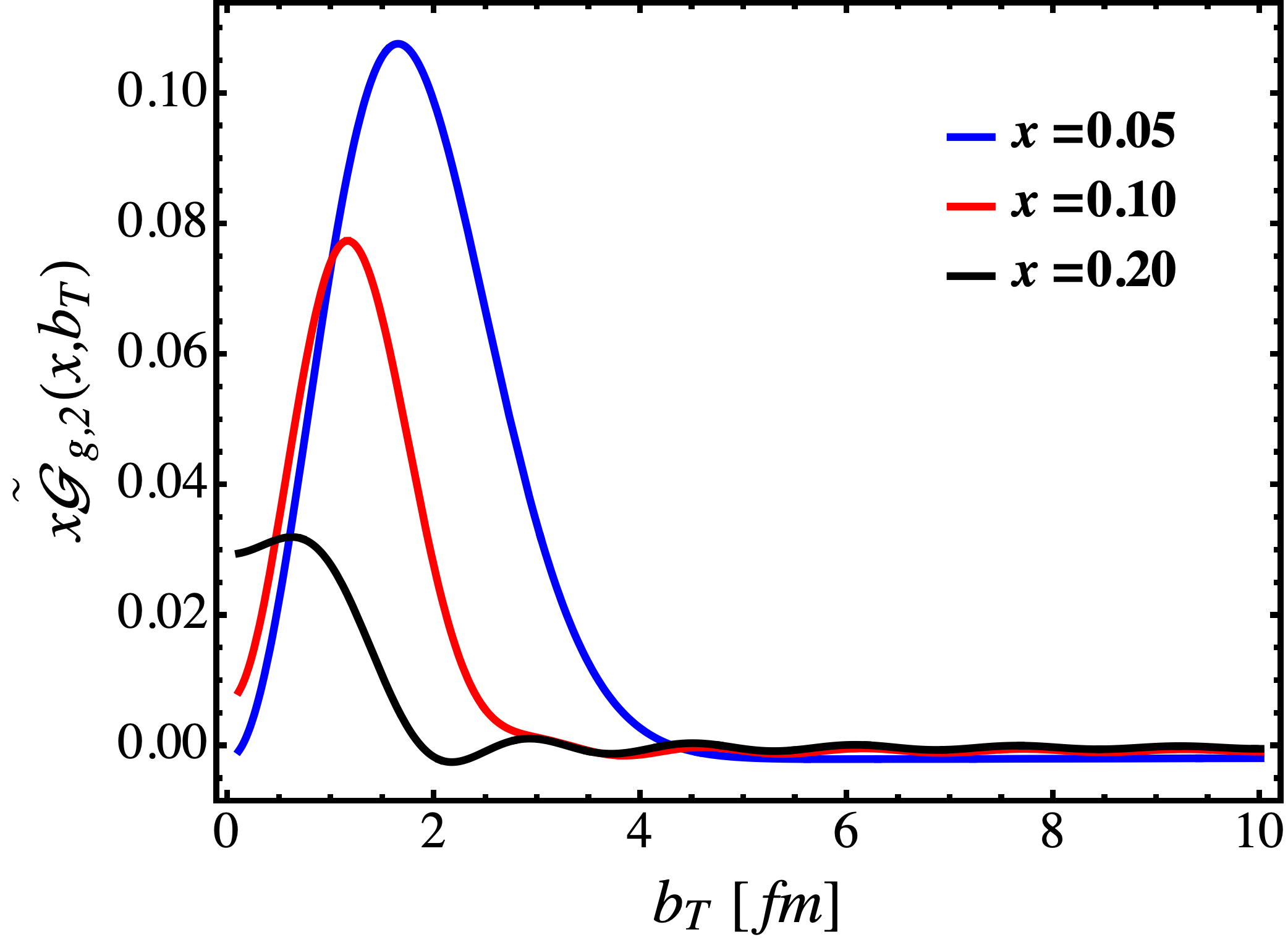} 
        \caption{}
    \end{subfigure}
    \hspace{1cm}
    \begin{subfigure}[b]{0.45\textwidth}
        \centering
        \includegraphics[width=\textwidth]{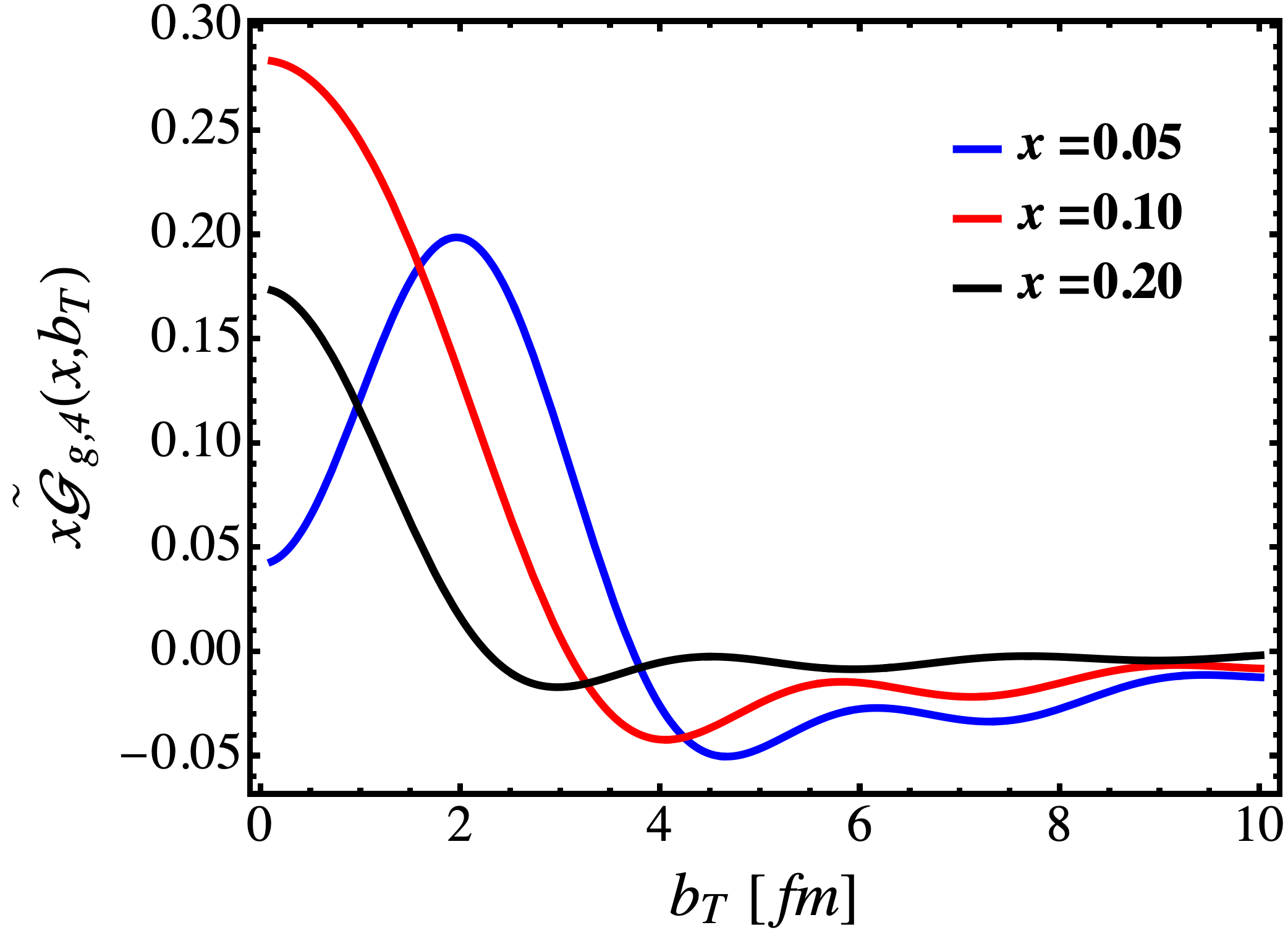} 
        \caption{}
    \end{subfigure}
    \caption{The twist-3 IPDPDFs \(x \mathcal{G}_{g,3}(x,b_T)\),\(x\mathcal{\tilde{G}}_{g,1}(x,b_T)\),\(x\mathcal{\tilde{G}}_{g,2}(x,b_T)\) and \(x\mathcal{\tilde{G}}_{g,4}(x,b_T)\) are plotted with respect to \(b_T[fm]\) in the kinematic range \(b_T\in[0.01,10]\) at fixed of \(x=0.05(\text{blue}),0.10(\text{red})\) and \(x=0.20(\text{black})\).}
    \label{Figure4}
\end{figure*}

\begin{figure*}[htbp]
    \centering
    \begin{subfigure}[b]{0.45\textwidth}
        \centering
        \includegraphics[width=\textwidth]{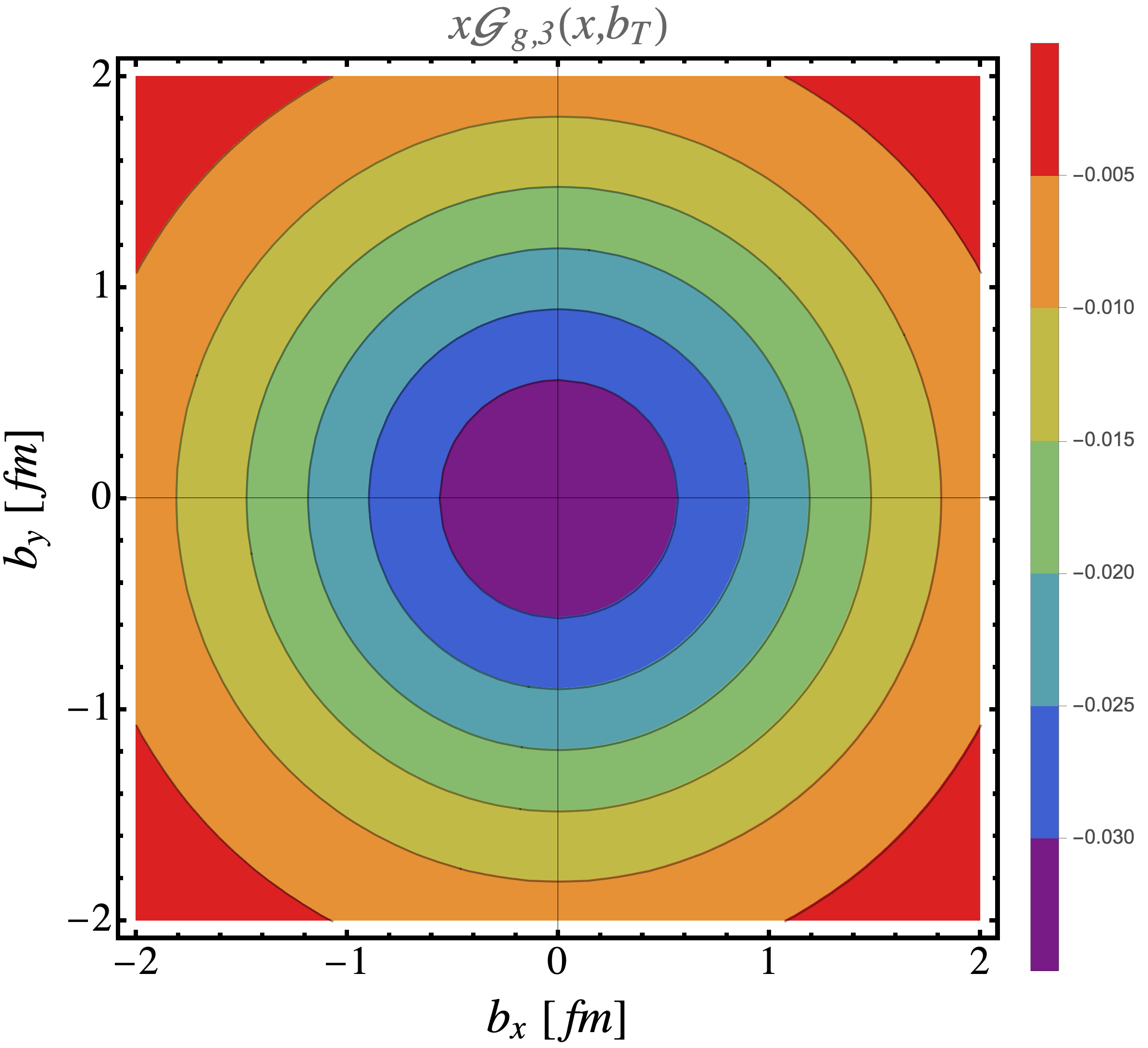} 
        \caption{}
    \end{subfigure}
    \hspace{1cm}
    \begin{subfigure}[b]{0.45\textwidth}
        \centering
        \includegraphics[width=\textwidth]{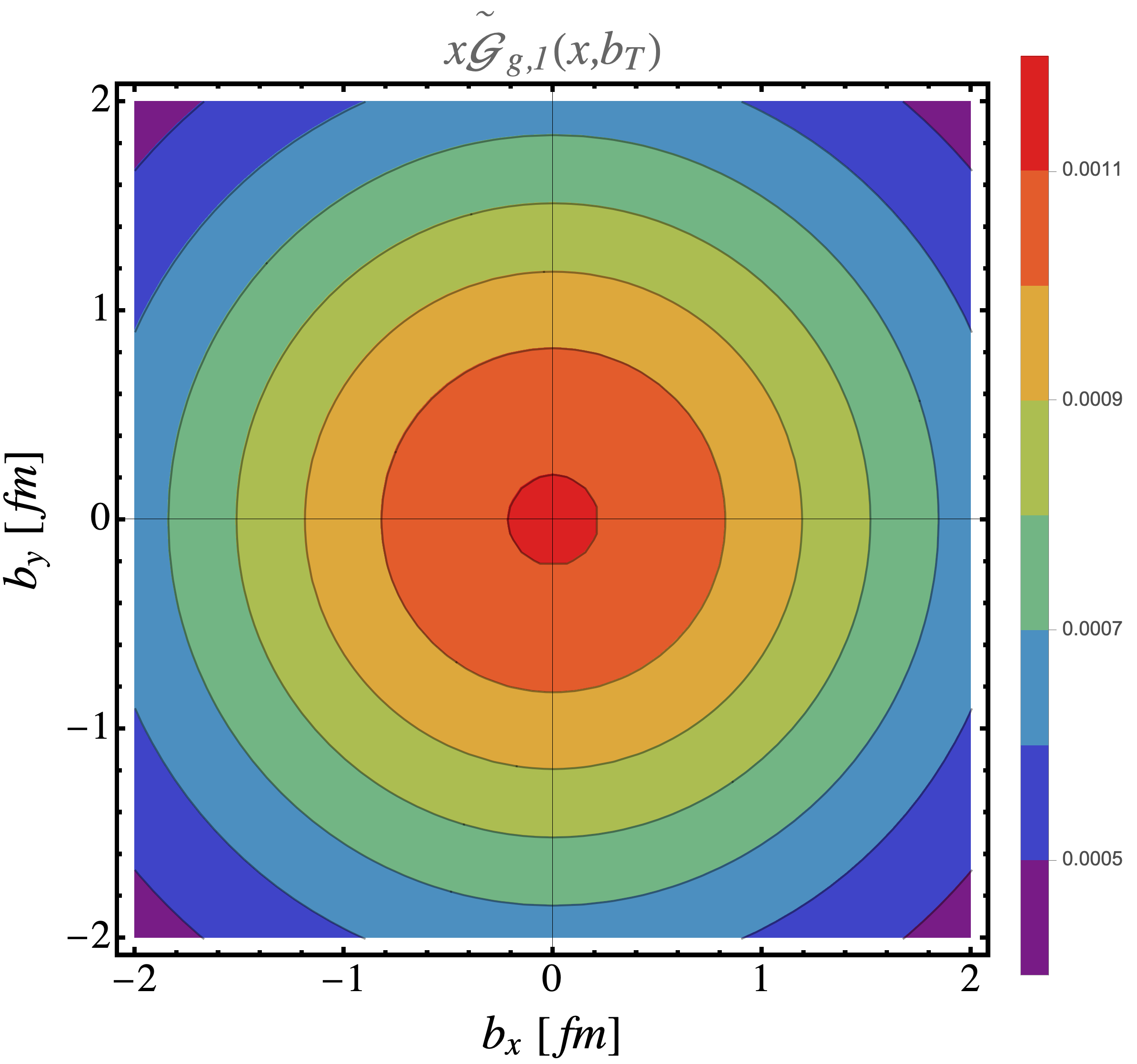} 
        \caption{}
    \end{subfigure}

    \vspace{1em}

    \begin{subfigure}[b]{0.45\textwidth}
        \centering
        \includegraphics[width=\textwidth]{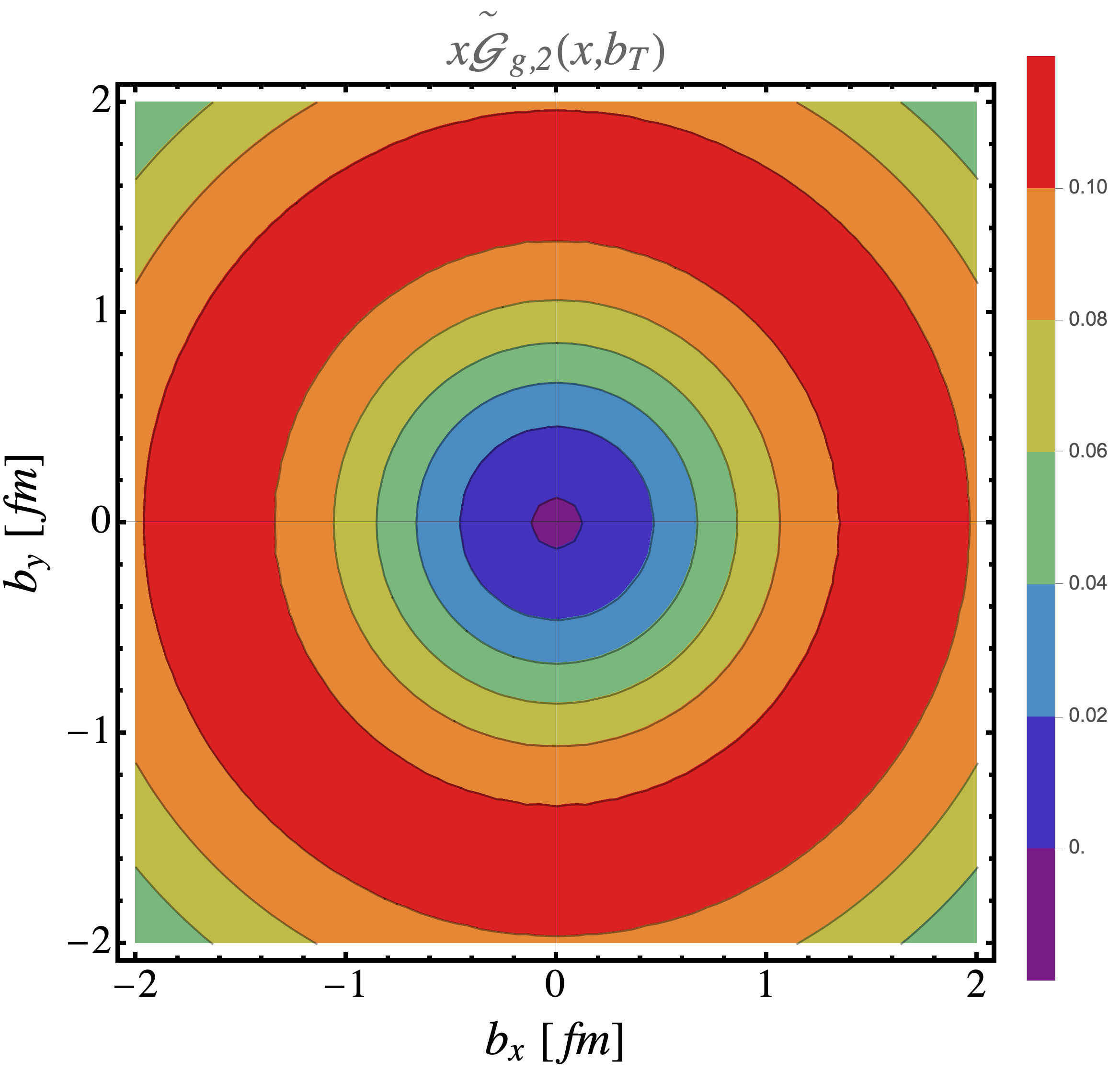} 
        \caption{}
    \end{subfigure}
    \hspace{1cm}
    \begin{subfigure}[b]{0.45\textwidth}
        \centering
        \includegraphics[width=\textwidth]{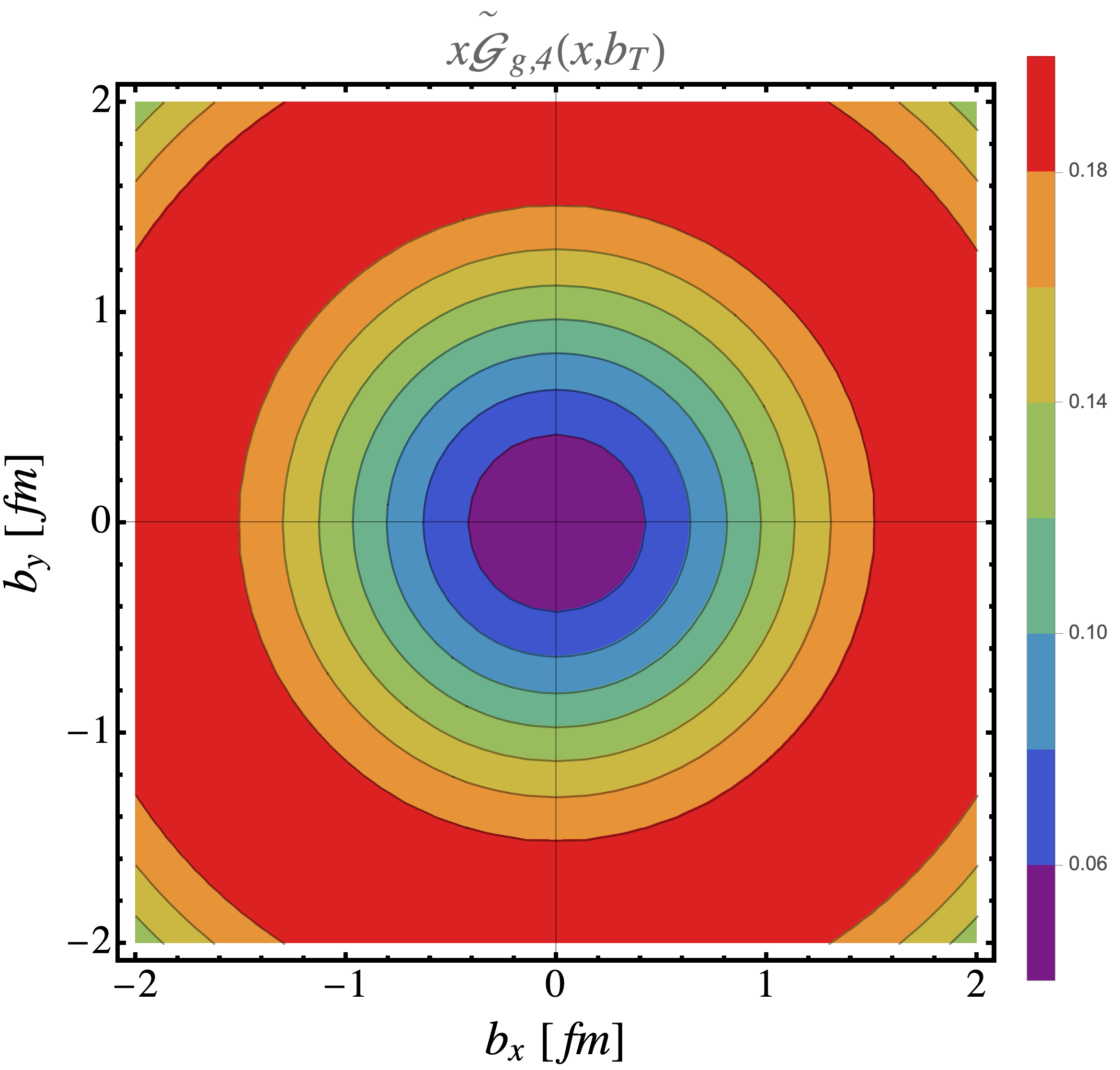} 
        \caption{}
    \end{subfigure}
    \caption{The twist-3 IPDPDFs \(x \mathcal{G}_{g,3}(x,b_T)\),\(x\mathcal{\tilde{G}}_{g,1}(x,b_T)\),\(x\mathcal{\tilde{G}}_{g,2}(x,b_T)\) and \(x\mathcal{\tilde{G}}_{g,4}(x,b_T)\) are plotted with respect to \(b_x\) and \(b_y\) at fixed value of \(x=0.05\).}
    \label{Figure5}
\end{figure*}

To gain further insight into the IPDPDFs, we present the two-dimensional projections of them in figure~\ref{Figure4}. In Fig.~\ref{Figure4}(a), the distribution \(x\mathcal{G}_{g,3}(x,b_T)\) is found to be negative, with a maximum peak at \(x=0.05\), followed by a sharper fall compared to \(x=0.10\). As \(x\) increases, the amplitude approaches zero, with all curves converging around \(b_T \approx 4~\text{fm}\). The distributions \(x\mathcal{\tilde{G}}_{g,1}(x,b_T)\) and \(x\mathcal{\tilde{G}}_{g,2}(x,b_T)\) remain positive over the full kinematic range \(b_T \in [0.01,10]~\text{fm}\). The function \(x\mathcal{\tilde{G}}_{g,1}(x,b_T)\) exhibits its highest peak at \(x=0.20\), although it decreases more rapidly than at smaller \(x\). For \(x=0.05\), it develops a bell-shaped peak around \(b_T \approx 2~\text{fm}\), which diminishes and shifts toward smaller values as \(x\) increases. Among all distributions, \(x\mathcal{\tilde{G}}_{g,4}(x,b_T)\) gives the largest contribution, with its maximum amplitude at \(x=0.10\). While positive at small \(b_T\), it changes sign beyond \(b_T \approx 2~\text{fm}\) before vanishing asymptotically.

To illustrate the spatial distribution of gluons, we present the impact parameter dependent distributions in Figs.~1--4 as functions of the transverse coordinates $b_x$ and $b_y$. The circular symmetry observed in these plots arises from the equal contribution of $b_x$ and $b_y$ in the unpolarized case, where the dependence enters only through $b_T=\sqrt{b_x^2+b_y^2}$. In situations where specific transverse directions are emphasized, such as in polarized distributions or spin densities, this symmetry can be broken, leading to azimuthally distorted (noncircular) profiles.

\subsection{Kinetic orbital angular momentum}

The gauge-invariant energy-momentum tensor (EMT) for gluons plays a crucial role in understanding the internal dynamics of hadrons in quantum chromodynamics (QCD). The gluon part of the EMT is defined as~\cite{Ji:1996ek}:

\begin{align}
    M^{\mu\nu}_g(\eta) = 2\, \mathrm{Tr}\left\{
        - G^{\mu\rho}(\eta) G^{\nu}_{\ \rho}(\eta)
        + \frac{1}{4}g^{\mu\nu} \left[ - G^{\rho\sigma}(\eta) G_{\rho\sigma}(\eta) \right]
    \right\},
    \label{eq:gluonEMT}
\end{align}
where $G^{\mu\nu}$ denotes the gluon field strength tensor, $g^{\mu\nu}$ is the Minkowski metric, $\eta$ denotes the space-time point, and $\mathrm{Tr}$ stands for the color trace.

To extract the angular momentum (AM) carried by gluons, it is necessary to focus on the transverse matrix elements of Eq.~\ref{eq:gluonEMT}. In the forward limit and considering nucleon matrix elements relevant for angular momentum, terms proportional to $g^{\mu\nu}$ do not contribute and can be omitted. The relevant correlator for the gluon part of the EMT on the light front (LF) can be written as:
\begin{align}
    M_g^{+T,\mathrm{LF}}(x)
    = -\int\frac{d\lambda}{2\pi}\, e^{i\lambda x}\, 2\, \mathrm{Tr}\left\{
        G^{+\eta}\left(-\frac{\lambda n}{2}\right)\,
        \mathcal{W}_{-\lambda/2,\, \lambda/2}\,
        {G}^{T}_\eta\left(\frac{\lambda n}{2}\right)
    \right\},
    \label{eq:LFcorrelator}
\end{align}

The local limit of this operator is obtained by integrating over $x$:
\begin{align}
    M_g^{+T,\,\mathrm{LF}}(0) = \int dx \, M_g^{+T,\,\mathrm{LF}}(x)
\end{align}

These correlators are directly connected to the twist-3 gluon generalized parton distributions (GPDs) introduced in Eq.~\ref{GPDs1}. Their moments are constrained by sum rules:
\begin{align}
    \int dx\, x\, G_{g,3}(x)\, &=\, A_g(0) + B_g(0), \\
    \int dx\, x\, G_{j,3}(x)\, &=\, 0,\quad \text{for} \quad j=1,2,4,
\end{align}
where $A_g(t)$ and $B_g(t)$ are the gluon gravitational form factors, evaluated at zero momentum transfer ($t=0$).

\begin{figure*}[htbp]
    \centering
    \begin{subfigure}[b]{0.45\textwidth}
        \centering
        \includegraphics[width=\textwidth]{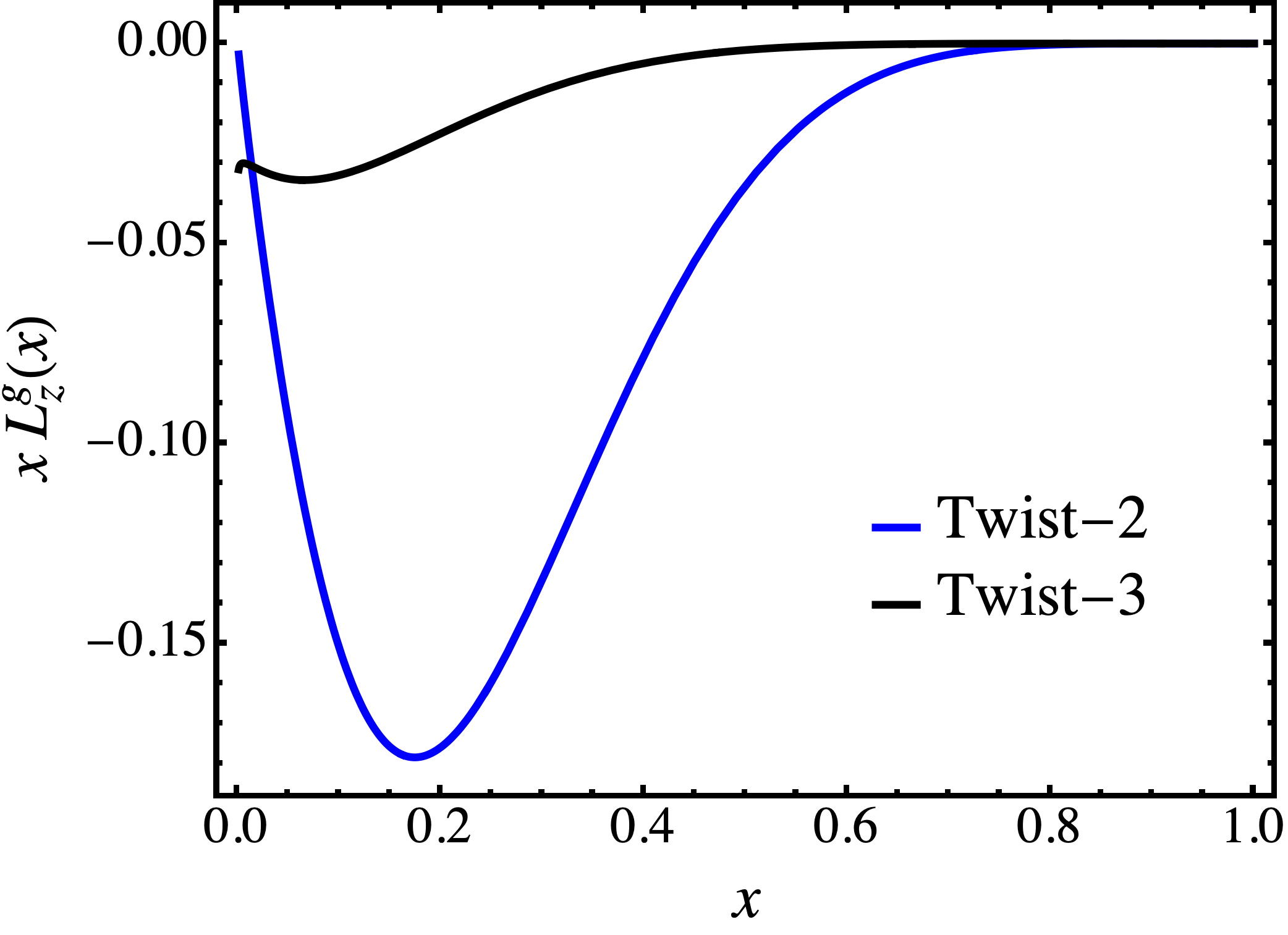} 
        \caption{}
    \end{subfigure}
    \hspace{1cm}
    \begin{subfigure}[b]{0.45\textwidth}
        \centering
        \includegraphics[width=\textwidth]{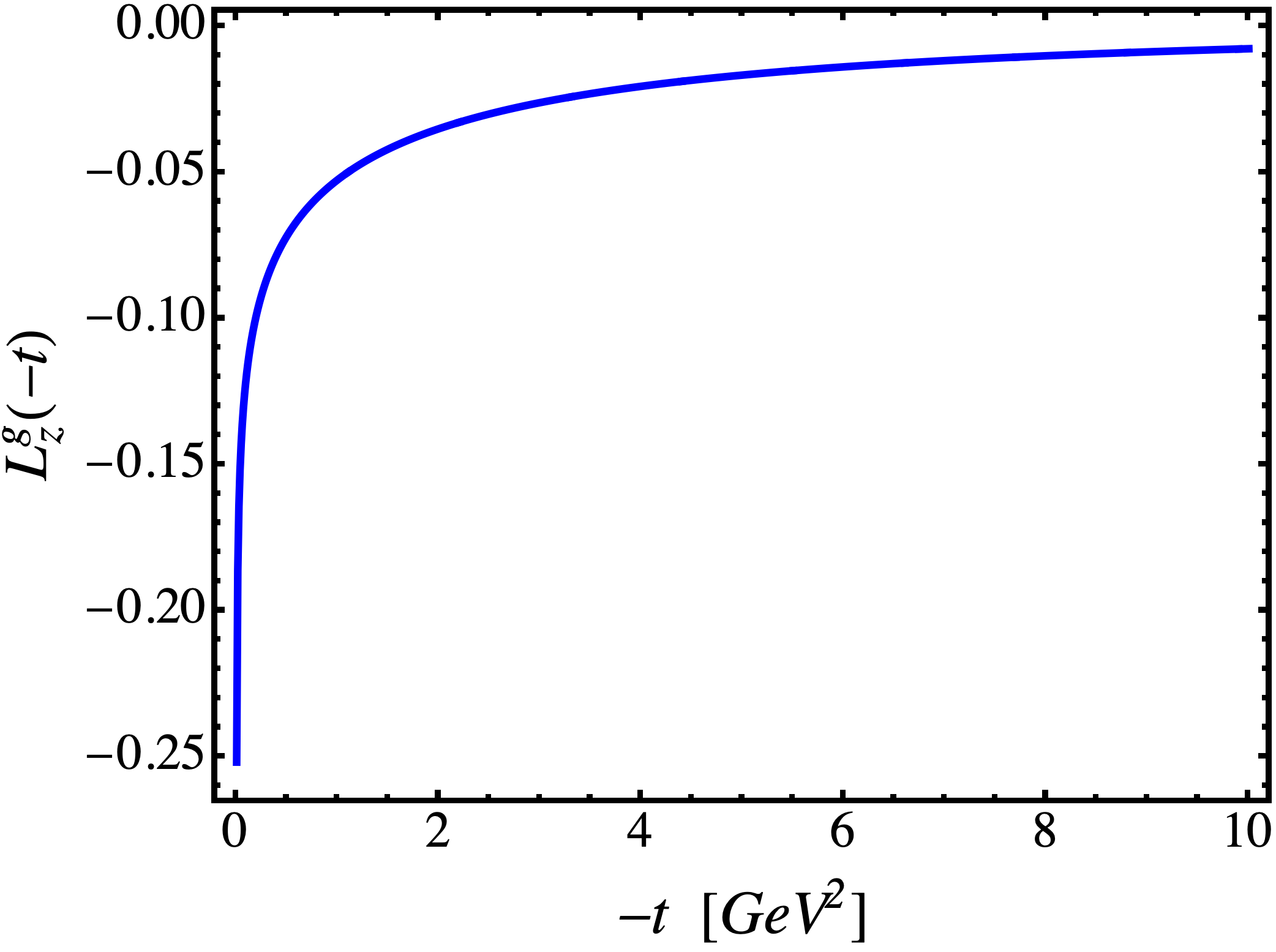} 
        \caption{}
    \end{subfigure}
    \caption{The unintegrated kinetic gluon OAM, $xL_z^g(x)$, as a function of $x$ is shown in panel (a), where the blue line corresponds to the twist-2 contribution and the black line to the twist-3 contribution. In panel (b), the kinetic OAM of gluon, $L_z^g(-t)$, is plotted as a function of the momentum transfer squared, $-t\,[\text{GeV}^2]$, in the range $-t \in [0,10]$.}
    \label{Figure6}
\end{figure*}
    
Consequently, the kinetic angular momentum carried by gluons in a longitudinally polarized nucleon \cite{guo2021novel} is given by:
\begin{align}
     L^g_z = \frac{1}{2} \left[ A_g(0) + B_g(0) \right]
     = \frac{1}{2}\int dx\, x\, G_{g,3}(x)
     \label{eq:Jg}
\end{align}

This fundamental relation establishes a direct connection between the first moment of the gluon GPD $G_{g,3}(x)$ and the total gluon angular momentum, encapsulated in the measurable gravitational form factors $A_g$ and $B_g$. 
The kinetic OAM is related to the leading order GPDs(twist-2) through Ji’s sum rule as follows \cite{ji1997gauge,LEADER2014163,wakamatsu2010gauge,chen2008spin}:

\begin{equation}
    L^g_z=\int dx \left[\frac{1}{2}x\left\{H^g(x,0,0)+E^g(x,0,0)\right\}-\tilde{H}^g(x,0,0)\right]
\end{equation}

In Fig.~\ref{Figure6}(a), we present the unintegrated distribution \(xL_z^g\) from both twist-2 and twist-3 contributions as a function of the gluon longitudinal momentum fraction \(x\). We observe that the twist-2 contribution exhibits a larger amplitude compared to the twist-3 part. While the twist-2 distribution vanishes in the small-\(x\) limit, the twist-3 contribution instead saturates to a constant value of approximately \(-0.04\). Figure~\ref{Figure6}(b) shows the dependence of \(L_z^g(-t)\) on the momentum transfer squared, \(-t\) [GeV\(^2\)]. From the numerical evaluation at \(-t\to0\), we obtain \(L_z^g=-0.252\), indicating that the gluonic kinetic OAM is negative.

For comparison, in the leading-twist sector, light-front model~\cite{Chakrabarti:2024hwx} reported a value of \(L_z^g=-0.42\), while another light-cone model calculation yielded \(L_z^g=-0.123\) \cite{PhysRevD.108.054038}. Thus, in this work, we provide a comparative analysis of the gluon angular momentum obtained from the twist-3 sector with that of the well-studied twist-2 contributions, and find good consistency with the available results.

\section{Conclusion}
\label{Sec:Conclusion}

In this work, we have investigated the twist-3 generalized parton distributions (GPDs) of gluons within the light-front model. The twist-3 GPDs were expressed in terms of the overlap representation of light-cone wave functions and evaluated numerically. We presented the behavior of the GPDs as functions of the longitudinal momentum fraction \(x\) and the transverse momentum transfer \(\Delta^2_T\), along with two-dimensional plots at fixed values of \(\Delta^2_T\) to provide additional insight into their structure. Furthermore, we explored the GPDs in impact-parameter space (IPDPDFs), obtained through a Fourier transform of the transverse momentum transfer. Our analysis shows that all IPDPDFs vanish at large impact parameter \(b_T\), indicating that gluons are confined within a finite region inside the hadron. In addition, we examined the twist-3 contributions to the kinetic orbital angular momentum (OAM) of gluons and validated our results by comparing them with known leading-twist calculations available in the literature. Importantly, to the best of our knowledge, this work presents the first systematic theoretical calculation of twist-3 gluon GPDs and their associated kinetic OAM within a light-front framework. No prior theoretical models, lattice QCD computations, or experimental analyses exist for these quantities at twist-3.

\bibliographystyle{JHEP}
\bibliography{biblio}

\end{document}